%% file: bare_jrnl.tex
\documentclass[10pt,journal, compsoc]{IEEEtran}
\usepackage{cite}
\usepackage{amsmath,amssymb,amsfonts}
\usepackage{algorithmic}
\usepackage{graphicx}
\usepackage{textcomp}
\usepackage{xcolor}
\usepackage{subfiles}
\usepackage{enumitem}
\usepackage{color, soul}

\ifCLASSINFOpdf
\else
\fi
\begin{document}
%
\title{Automated Machine Learning for \\ Deep Learning based Malware Detection}
%
%

\author{Austin Brown, Maanak Gupta,~\IEEEmembership{Senior Member,~IEEE}, and
        Mahmoud Abdelsalam
        }

%
%

\markboth{Brown et al.: Automated Machine Learning for Deep Learning based Malware Detection}%
{Shell \MakeLowercase{\textit{et al.}}: Bare Demo of IEEEtran.cls for IEEE Journals}
%




\IEEEtitleabstractindextext{
\begin{abstract}
Deep learning (DL) has proven to be effective in detecting sophisticated malware that is constantly evolving. Even though deep learning has alleviated the feature engineering problem, finding the most optimal DL model's architecture and set of hyper-parameters, remains a challenge that requires domain expertise. In addition, many of the proposed state-of-the-art models are very complex and may not be the best fit for different datasets. A promising approach, known as Automated Machine Learning (AutoML), can reduce the domain expertise required to develop custom DL models by automating the ML pipeline key components, namely hyperparameter optimization and neural architecture search (NAS). AutoML reduces the amount of human trial-and-error involved in designing DL models, and in more recent implementations can find new model architectures with relatively low computational overhead.

Research on the feasibility of using AutoML for malware detection is very limited. This work provides a comprehensive analysis and insights on using AutoML for both static and online malware detection. For static, our analysis is performed on two widely used malware datasets: SOREL-20M to demonstrate efficacy on large datasets; and EMBER-2018, a smaller dataset specifically curated to hinder the performance of machine learning models. In addition, we show the effects of tuning the NAS process parameters on finding a more optimal malware detection model on these static analysis datasets. Further, we also demonstrate that AutoML is performant in online malware detection scenarios using Convolutional Neural Networks (CNNs) for cloud IaaS. We compare an AutoML technique to six existing state-of-the-art CNNs using a newly generated online malware dataset with and without other applications running in the background during malware execution. We show that the AutoML technique is more performant than the state-of-the-art CNNs with little overhead in finding the architecture. In general, our experimental results show that the performance of AutoML based static and online malware detection models are on par or even better than state-of-the-art models or hand-designed models presented in literature.
\end{abstract}

}

\maketitle

%
\IEEEpeerreviewmaketitle

\subfile{sections/Introduction}
\subfile{sections/Background.tex}

\subfile{sections/Static_AutoML.tex}

\subfile{sections/Online_AutoML.tex}

\subfile{sections/Future_Work_Conclusion.tex}

\bibliographystyle{IEEEtran}
\bibliography{references}

%

\begin{IEEEbiography}[{\includegraphics[width=1.1in,height=1.33in]{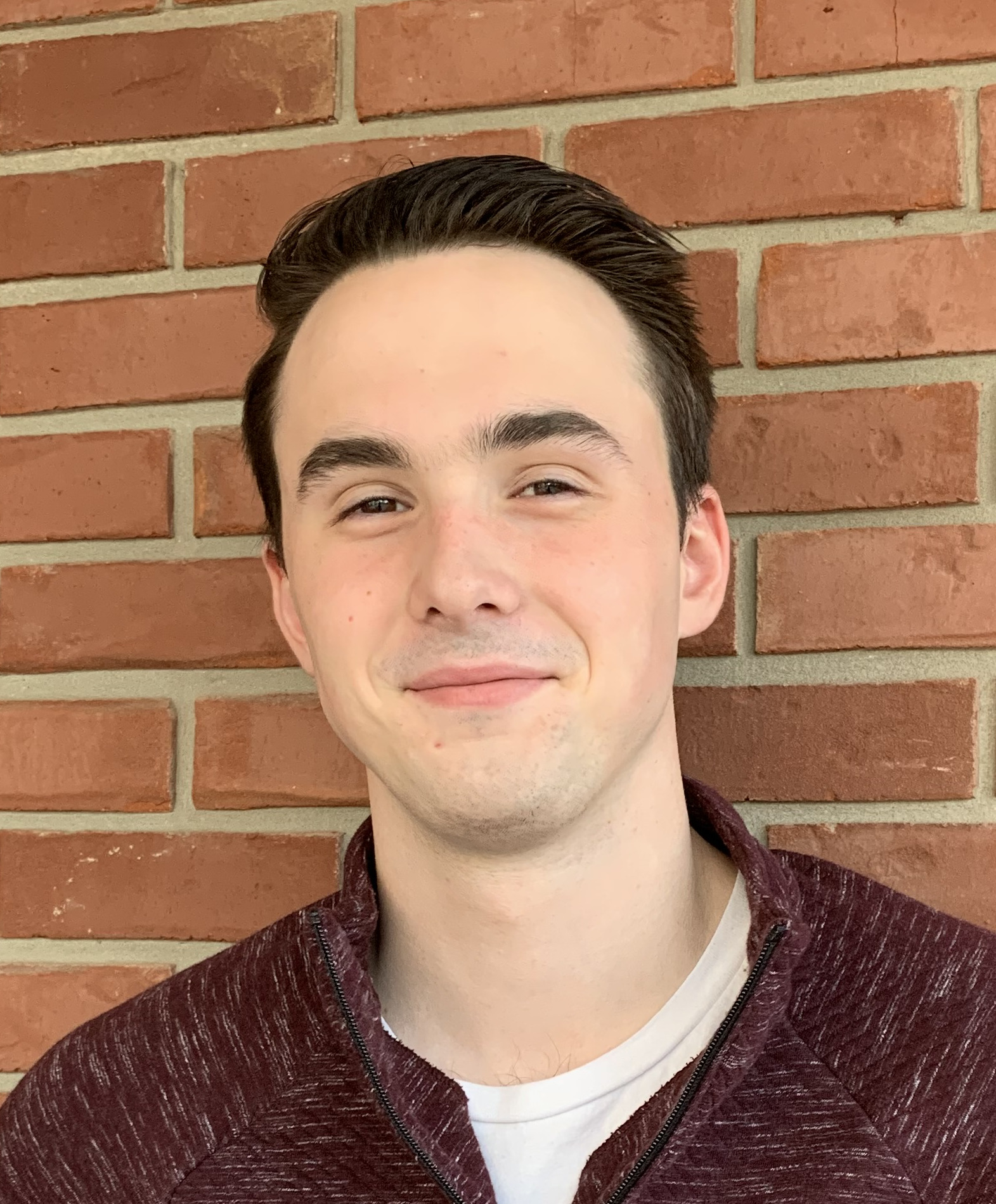}}]{Austin Brown}
Received his B.S. in Computer Science from Tennessee Tech University in 2020. He received his M.S in Computer Science from Tennessee Tech University in 2022. His interests include deep learning, malware research, and cloud computing. 
\end{IEEEbiography}



\begin{IEEEbiography}[{\includegraphics[width=1.1in,height=1.33in]{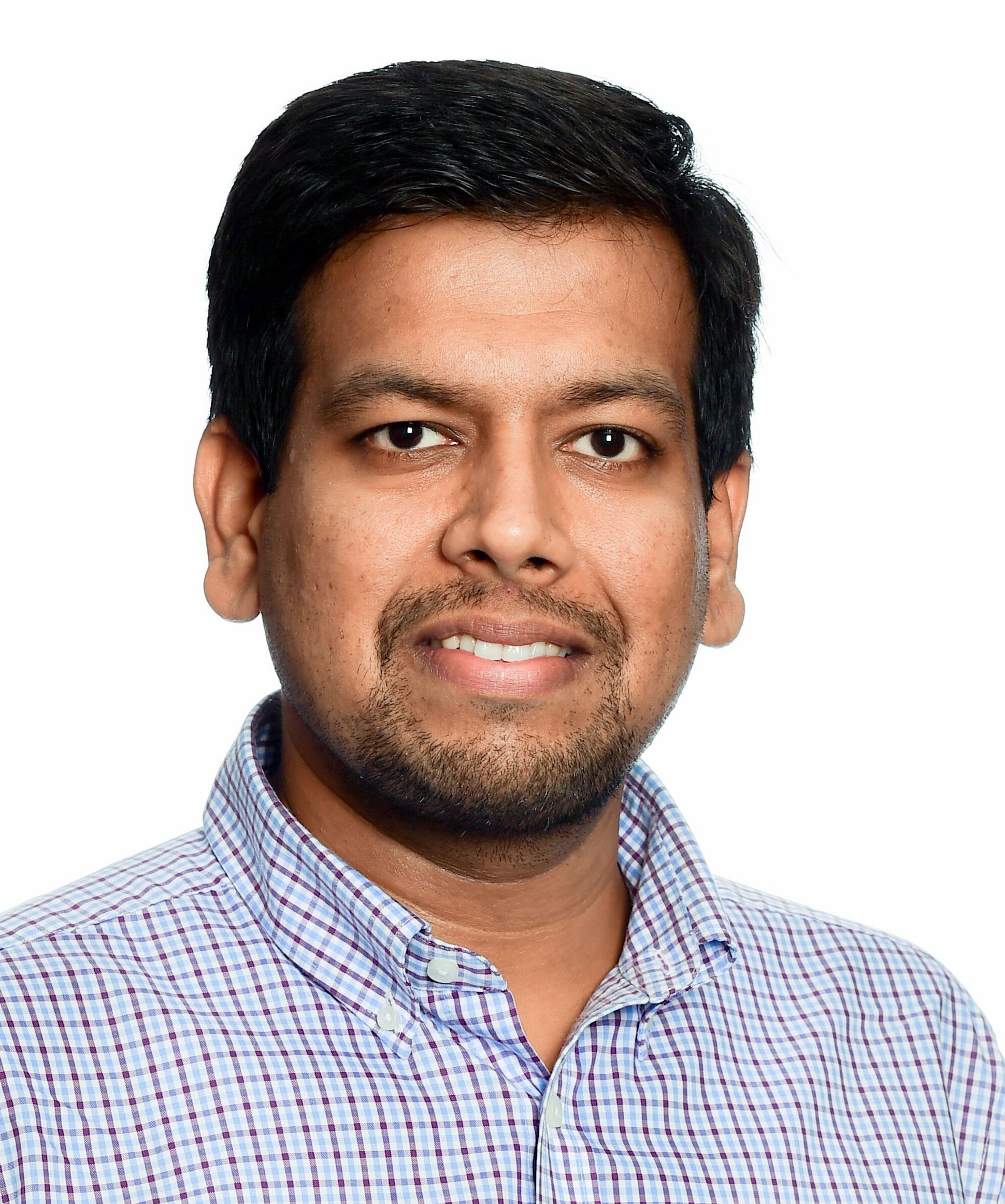}}]{Maanak Gupta}
is an Assistant Professor in Computer Science at Tennessee Tech University, Cookeville, USA. He received M.S. and Ph.D. in Computer Science from the University of Texas at San Antonio (UTSA) and has also worked as a postdoctoral fellow at the Institute for Cyber Security (ICS) at UTSA. His primary area of research includes security and privacy in cyber space focused in studying foundational aspects of access control, malware analysis, AI and machine learning assisted cyber security, adversarial AI and their applications in technologies including cyber physical systems, cloud computing, IoT and Big Data.  He holds a B.Tech degree in Computer Science and Engineering from Kuruskhetra University, India, and an M.S. in Information Systems from Northeastern University, Boston. He is senior member of IEEE.

\end{IEEEbiography}

\begin{IEEEbiography}[{\includegraphics[width=1.1in,height=1.3in,clip,keepaspectratio]{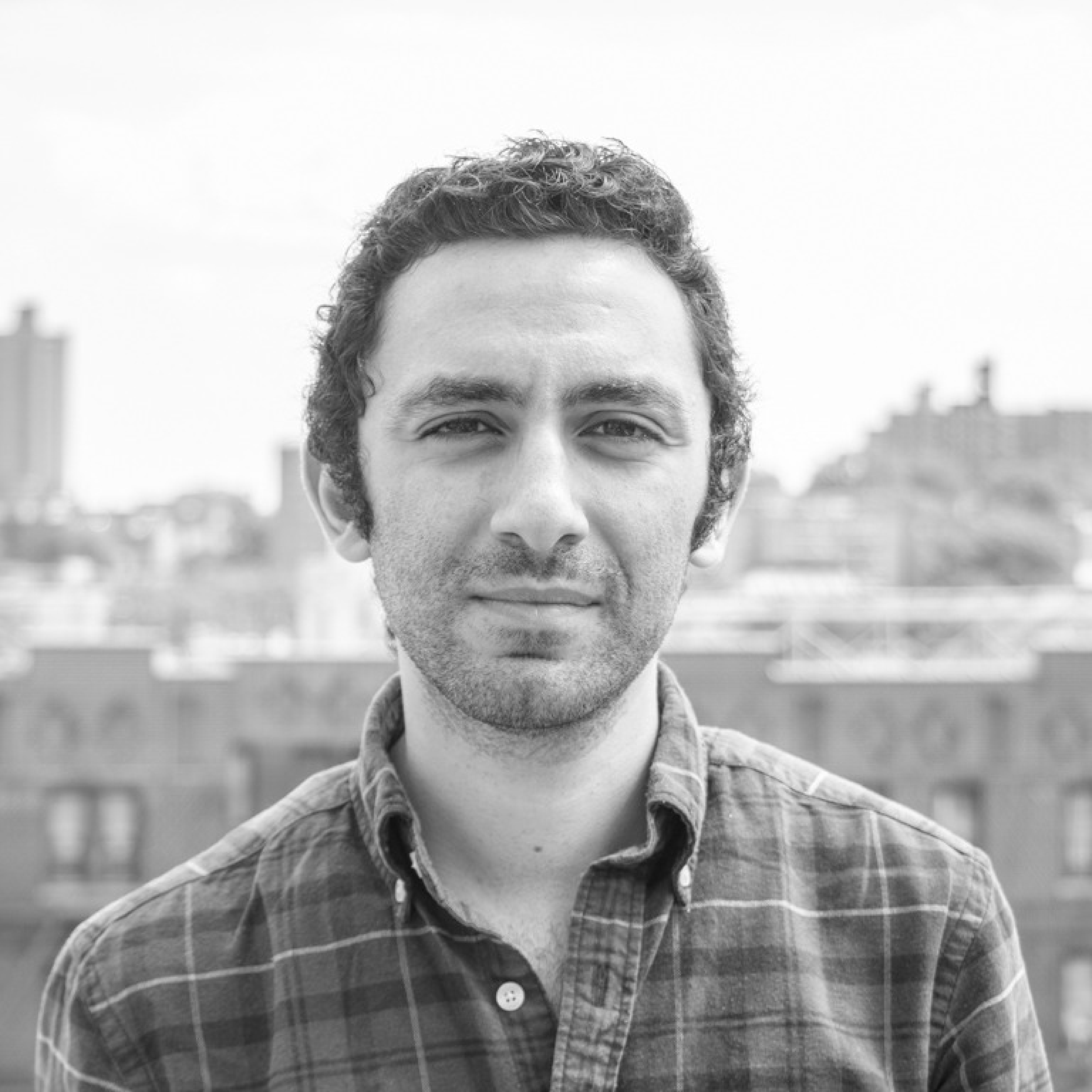}}]{Mahmoud Abdelsalam}
received the B.Sc. degree from the Arab Academy for Science and Technology and Maritime Transportation (AASTMT), in 2013, and the M.Sc. and Ph.D. degrees from the University of Texas at San Antonio (UTSA), in 2017 and 2018, respectively. He was working as a Postdoctoral Research Fellow with the Institute for Cyber Security (ICS), UTSA, and as an Assistant Professor with the Department of Computer Science, Manhattan College. He is currently working as an Assistant Professor with the Department of Computer Science, North Carolina A\&T State University. His research interests include computer systems security, anomaly and malware detection, cloud computing security and monitoring, CPS security, and applied machine learning.
\end{IEEEbiography}




\end{document}

%% file: sections/Introduction.tex
\section{Introduction}\label{sec:intro}

\subsection{Overview and Motivation}

Malware is becoming a more profitable domain for malicious actors with the rise of digital connectivity and the growing critical infrastructures reliance. These cyberattacks have costed the industry billions \cite{anderson2019measuring_cyber_crime_cost} of dollars. The increase and impact of such cyberattacks has called for novel and sophisticated defense mechanisms in response to those that wish to protect digital assets from malware attacks.

There are several existing approaches for malware analysis, including static \cite{nath2014static, shalaginov2018machine}, dynamic \cite{willems2007toward_dynamic, tobiyama2016malware,alotaibi2019identifyingmalware}, and online analysis \cite{mcdole2021deep, mcdole2020analyzing_cnn, kimmel2021recurrent_rnn, kimmell2021analyzing}. Each of these analysis methods collect different features from the file or system in question, ranging from details of the file header in static analysis, to holistic operating system level performance metrics in the case of online analysis. The reasons to use a specific analysis approach depends on the use case and availability of data. For simple file scanning, static analysis is the fastest method, since there is no need to run the executable in question, whereas, collecting data from a running executable in dynamic analysis may give more insight into the true behavior and intent of a questionable executable. On the other hand, unlike dynamic and online analysis, static analysis can be crippled using well-known obfuscation techniques.

Machine learning (ML), especially DL, has become a popular technique to develop malware detection solutions, and has shown promising results \cite{sahin2020survey} because of its ability to learn generalized patterns to identify unseen malware. As such, research works \cite{kimmell2021analyzing, xie2020system,gupta2023chatgpt, xiao2019android, mishra2019cloud, tobiyama2016malware, kolosnjaji2016deep, pascanu2015malware,aryal2021survey, brown2022online, aryal2022analysis} have employed different types of ML models to detect malware on a variety of systems and data sources, depending on the use case.
These proposed solutions have utilized both traditional ML algorithms and, more recently, deep learning algorithms.
Approaches \cite{n-gram-mal-code, fan2016maliciousmalware} that rely on traditional machine learning models require domain experts for feature engineering, which, in most cases, is burdensome and laborious. On the contrary, deep learning based approaches \cite{deepmalnet_journal, deep_mal_detect_Sewak, malconv, rudd2019aloha,ganesh2017cnn, wang2019effective_cnn, yeo2018flow_cnn, kimmel2021recurrent_rnn, agrawal2019attention_rnn, jha2020recurrent_rnn, luckett2016neural} eliminate the feature engineering step and are gaining more traction.
Some works \cite{mcdole2020analyzing_cnn, rezende2018malicious_vgg} have utilized state-of-the-art DL models (e.g., ResNet\cite{he2016deep_resnetpaper}, DenseNet\cite{iandola2014densenet}, and VGG16\cite{simonyan2014very_vgg_paper}) that perform well in general and train it on malware data; however, these models are usually very complex and require a rigorous tuning process to achieve the desired high performance. In addition, such models are usually designed for tasks like image, text, or voice recognition and can be inadequate for malware detection. Consequentially, works \cite{malconv} have focused on manually crafting model architectures that fit the malware detection domain. However, these approaches not only require heavy tuning, but also high technical skills in both the ML and the malware domains.

Automated Machine Learning (AutoML) \cite{he2021automl} seeks to automate the process of finding an optimal model architecture for the given data and tuning this model to achieve higher performance. In addition, it can also reduce the work needed to redesign a malware detection model as malware and data sources evolve overtime. Even though AutoML pipelines require more computational time to produce a model, they significantly reduce the work hours and expertise needed to find a performant model.

AutoML holds significant promise for malware detection, automating the dual tasks of discovering the ideal machine learning architecture tailored for this purpose and subsequently refining this selected model. Despite its potential, AutoML is still in its developmental phase, and comprehensive exploration of its applicability, especially in malware detection, remains limited.

Building on this, AutoML serves as a potent tool for domain experts, even those without the formal "Data Scientist" designation, enabling them to harness machine learning more effectively. The efficacy of AutoML, like many machine learning realms, often hinges on the volume of accessible data and the intricacy of the given task. Domains rich in data, such as our focus on malware detection, as well as simulated environments that can generate vast amounts of data, are well-poised to reap the benefits of AutoML. While our discussion centers on its utility in static and online malware detection, AutoML's potential can be extrapolated to other cybersecurity areas like Network Intrusion Detection, Security Log Analysis, and Threat Intelligence. Essentially, any domain with ample high-quality data suited for deep learning might find AutoML advantageous.






With the growth in malware sophistication and machine learning complexity, especially in deep learning, finding the most performant deep neural architecture without a significant increase in human hours spent is critical. This paper aims to study the feasibility of integrating AutoML into the malware detection pipeline to remove the need to hand design and tune ML models. In particular, we focus on using deep learning, specifically Feed Forward Neural Networks (FFNNs) and Convolutional Neural Networks (CNNs). FFNNs have a high level of expressive power and require much less feature engineering than traditional machine learning approaches. Convolutional Neural Networks can model complex functions with image shaped data as input with little to no feature engineering, only requiring framing the data in a 2d vector. Further, we focus on both data that is gathered through static analysis, specifically focusing on portable executable (PE) files which are the predominant executable format in the Windows operating system, and online data captured from running, internet connected, Linux servers in cloud IaaS with malware executed on them.
The \textit{main contributions} of this work are:
\begin{itemize}
    \item We study the feasibility of using AutoML for deep learning based static malware detection and demonstrate the effectiveness of the produced AutoML Deep FFNN models by showing that they are comparable to manually crafted models, even without significantly tuning the AutoML pipeline.
    \item We provide insights and analysis of the automation parameters of the AutoML process on static malware data, and show how these parameters can affect the performance of the found optimal model. 
    \item We show that AutoML derived Convolutional Neural Networks can preform better than state-of-the-art Convolutional Neural Networks on online malware data, with little overhead in deriving the model architecture.
    \item We discuss ideas and future directions for improving the efficiency and performance of AutoML models that are designed for malware detection.
\end{itemize}



The rest of this work is organized as follows. Section \ref{sec:Background} discusses background and related works in this domain. Section \ref{sec:static_automl} shows the application of AutoML in two popular static malware datasets, with comparison to other works, and discussion of the presented AutoML methodology. Section \ref{sec:online_automl} focuses on one-shot AutoML applied to CNNs to detect malware in online cloud IaaS, with comparison to detection results of state-of-the-art CNNs on the same dataset. Section \ref{sec:future_work_conclusion} presents ideas for future work and improvements, as well as the conclusion to the findings in this work.

%% file: sections/Background.tex

\section{Background and Related Works}\label{sec:Background}


\subsection{Malware Detection}

\subsubsection{Static Analysis}
Static analysis involves analyzing features that can be observed in a binary without running the binary executable. Static analysis methods may include observations such as: file entropy; n-gram analysis of byte sequences in a binary; imports and API calls; strings found within the binary; header information. The major benefit of static analysis is its speed and low overhead, since the binary is not executed. 

One of the most simple forms of static analysis for malware detection is looking up the signature of a binary, most often the file hash. This method is extremely efficient if the binary's hash is documented, but has no ability to detect modified or new malware. A more popular method of static analysis looks at n-grams of bytes in the binary. Authors in \cite{n-gram-mal-code} measured frequency of common n-gram bytes in Windows binaries to determine if the binary is malicious. The frequency of n-grams across both malicious and benign binaries were used to train a K-nearest-neighbors classifier. While this approach showed good results (at the time published), it is unclear if it would show as good of results in modern malware detection. This approach additionally has proven to be computationally expensive and offers diminishing returns as \textit{n} increases \cite{raff2018investigation}. 
Another work \cite{fan2016maliciousmalware} has taken it a step further from n-gram byte analysis to analyzing instruction sequences in questionable binaries. 

To reduce the overhead imposed by the essential feature engineering in traditional ML, some researchers have focused on deep learning approaches. Authors in \cite{raff2017learning} used recurrent neural networks to analyze the first 300 bytes of the header of Windows PE files.
Work in \cite{malconv} utilizes convolutional layers within a neural network to extract information on Windows PE headers to determine binary intent.   
Authors in \cite{deepmalnet_journal} implemented what they call \textit{Windows-Static-Brain-Droid,} which implements multiple architectures in a voting scheme. The features for the architecture are both raw byte information and parsed features from the binary. The raw byte features feed into various architectures based on \cite{malconv}. The parsed features feed into multiple traditional classifiers and a FFNN. Section \ref{sec:static_automl} will focus on developing an optimized neural architecture similar to \cite{deepmalnet_journal}'s FFNN. 

\subsubsection{Dynamic Analysis}
Unlike static analysis, dynamic analysis executes a binary to monitor its behavior. This is most often carried out in a sandboxed environment to restrict the binaries access to other resources which a malware could attack. Data collected from the execution behavior of malware allows for greater insight into a questionable binary's intent and nature.
Authors in \cite{fan2016maliciousmalware, luckett2016neural} utilized system calls captured during execution to detect malware. Work in \cite{fan2016maliciousmalware} utilizes traditional machine learning approaches, while \cite{luckett2016neural} uses neural networks for classification. Authors in \cite{tobiyama2016malware}, look at API calls made in 5 minute intervals to classify binaries as benign or malicious. These calls were passed to a CNN for classification. In \cite{huang2016mtnet}, authors use FFFNs to classify binaries based on extracted API calls from dynamic execution. 

Compared to static analysis, these methodologies require extra computational overhead and time to detect malware. However, dynamic analysis will not be able to detect sophisticated malware that can detect the presence of an emulation sandbox or the lack of network connectivity that is often found with isolated emulation environments. 

\subsubsection{Online Analysis}
Where dynamic analysis only analyzes the execution of a single binary, online analysis collects data from an entire system to monitor (in real time) for malware execution. This allows for continuous monitoring of an open system (not in a sandbox), with full access to all resources. Additionally, this allows for collection of execution details that extend beyond that of a single binary. This can include both knowledge of normal execution of a given system as well has effects to adjacent processes from live malware execution.

The authors in \cite{demme2013feasibility, ozsoy2015malware} utilize performance counters from an entire system to detect the presence of malware. Guan \textit{et al.} \cite{guan2012ensemble} proposed using system calls to detect malware in online systems with ensembles of Bayesian predictors and decision trees. Others have proposed using memory features \cite{xu2017malware}. 
McDole et al. \cite{mcdole2021deep} and Abdelsalam et al. \cite{abdelsalam2019onlinemalware} show that per-process performance metrics from Ubuntu machines can provide high detection performance when ingested with a CNN. The process data is structured in the shape of an image, with the rows denoting different processes and the columns denoting different performance metrics for each process, collected from the target machine. Abdelsalam et al. achieves 89.5\% detection accuracy using shallow CNNs, while McDole et al. achieves 92.9\% detection accuracy using state-of-the-art CNNs on the same dataset.  Jeffery et al. \cite{kimmel2021recurrent_rnn} uses recurrent neural networks (RNNs) on the same dataset as McDole et al. and Abdelsalam et al., but organizes the inputs to the RNN as sequences of unique process features, all from the same time slice. They achieve 99.61\% detection accuracy with this technique. Online malware detection can incur high overhead with continuous monitoring of systems, but provides real-time detection performance on evasive and low-lying malware in a live environment without requiring the identification of a suspicious executable.

\subsection{Deep Learning for Malware Detection}
Using deep learning for malware detection has been researched extensively and  
spans approaches that utilize various types of deep learning algorithms including CNNs \cite{ganesh2017cnn, wang2019effective_cnn, yeo2018flow_cnn,  abdelsalam2018malwaremalwarecnn, mcdole2020analyzing_cnn}, RNNs  \cite{kimmel2021recurrent_rnn, agrawal2019attention_rnn, jha2020recurrent_rnn}, feed forward neural networks (FFNNs) \cite{deepmalnet_journal, deep_mal_detect_Sewak, rudd2019aloha}, etc.
Deep learning approaches presented in these works have the advantage over traditional ML models as they do not require hand designed features in order to be performant. 
Although these approaches impose additional performance overhead as compared to some traditional ML algorithms, many have shown to be more performant under some conditions \cite{kamath2018comparative}, with high accuracy in malware detection \cite{kimmell2021analyzing}. 

Despite the fact that deep learning approaches have shown tremendous results for malware detection, most of these works fall short because either (1) they utilize state-of-the-art models that are not tailored specifically for malware detection and may not be optimal for the data available, or (2) they have hand designed their models specifically for malware detection, without AutoML, which requires extensive domain experts' knowledge and hand tuning. Fortunately, AutoML can help overcome these obstacles and attain higher optimal results; however, the feasibility of utilizing AutoML for malware detection is hardly explored.

\subsection{AutoML Overview}

\subsubsection{Neural Architecture Search}

The performance of a model is highly dependent on the design of its architecture. A neural architecture search aims to find the architecture design that achieves the highest performance on unseen validation data. We consider a change in architecture design to constitute a change to the number or configuration of trainable parameters, or the layers' activation function. 

\subsubsection{One-Shot Search Methodology}

\begin{figure}[!t]
    \centering
        \includegraphics[width=.7\linewidth]{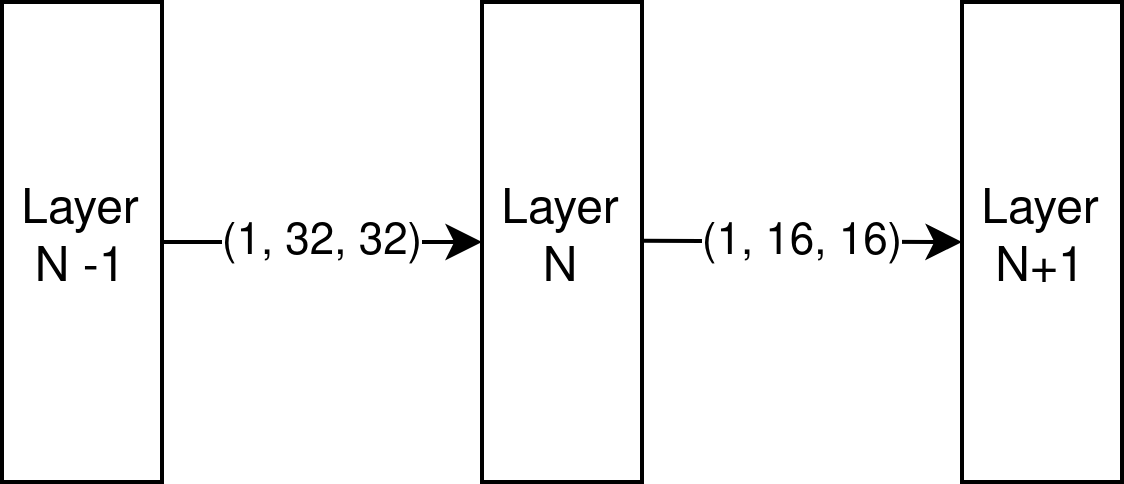}
        \caption{Example Convolutional Layer Output Shapes}
\label{fig:conv_layers}
\end{figure}

Many recent NAS methodologies focus on the computer vision domain. This field is heavily dependent on convolutional neural networks. Many types of layers within these networks, given the same shape of input, will produce the same shape of output. A network whose layers meet this condition is, intuitively, easily mutable; layers can be swapped out interchangeably, allowing the next layer to accept any chosen layer type's input since they share the same output shape as shown in Figure \ref{fig:conv_layers}. Many types of layers can be substituted for \textit{Layer N} and maintain the same output shape of \textit{(1, 16, 16)}. This property allows for an algorithm to test multiple layer choices at each layer of a network to find the best architecture configuration. The work presented in \cite{pham2018efficient} can create a \textit{super-graph} which contains multiple \textit{sub-graphs} representing all permutations of networks given the choices of each layer. A similar work, \cite{liu2018darts} relaxes the constraints of the categorical layer choice to a softmax choice, such that the categorical choice is now continuous, and a gradient can be used to find the best layer choices through training by backpropagation.

These NAS methodologies are used to learn an entire network architecture or learn the architecture of a cell that is repeated throughout the network. As long as each of the operations (layer) choices produce the same shape of output, the specific operation choices within a cell can be anything. This possibility allows for designing not only convolutional cells, but also recurrent cells. This allows the algorithm to find both CNNs and RNNs, or a combination of both.

These algorithms, known as One-Shot algorithms, find the most performant network configuration in "one-shot", without the need to train the network from scratch multiple times, by leveraging theoutput shape property. 

\subsubsection{Multi-Trial Search Methodology}

Multi-Trial NAS solutions, as opposed to one-shot, require many trials of different network configurations to find a performant architecture. In the past, before the invention of clever one-shot methodologies, this was the only way to test out different network architectures. Today, some types of networks still rely on multi-trial NAS, such as networks that can't easily swap out layers because of layer output shapes. There has been some work to improve the efficiency of this process, such as \cite{wei2016network}, through weight sharing. This allows each trial to run for a much shorter amount of time by leveraging the learned parameters from previous trials, and only optimizing for new parameters. However, these algorithms, if not carefully controlled, can become unstable in later trials. For this reason, our work with deep feed forward neural networks in Section \ref{sec:static_automl} utilizes the more primitive multi-trial methodology in searching for the most performant architecture.

\subsubsection{NAS Search Space}
The NAS search space is the total space containing the values of all valid model configurations. During the NAS process, architecture configurations are drawn from this space and evaluated. The search space is arbitrarily large, so reasonable constraints are placed in order to bound the cost of time required to search and limit the complexity of the chosen model. For example, a model depth of 1,000 layers is a valid choice for an architecture, but it will produce a very complex model with a considerably high training time. For this reason, upper boundaries are usually provided. For example, in our proposed approach in Section \ref{sec:static_automl}, we set the number of layers' upper bound to 14 to limit the complexity of the model architectures available within the search space. Beside bounding the range of the search space, we also considered the sampling granularity and distribution of the search space. For example, in Section \ref{sec:static_automl} we set the granularity in selecting a layer's width to 128 neurons in order to limit the number of available selections while still maintaining an appropriate level of expression of its effect on model performance.
In order to simplify the NAS process, when a parameter value is selected from the search space, we fix this value throughout the model, instead of on a per-layer basis.

\subsubsection{Automated ML for Malware Detection}
Automated machine learning has recently been used in a variety of fields. Several AutoML works have been designed for specific domains, such as processes developed for the computer vision domain \cite{liu2018darts, pham2018efficient}. However, AutoML is still at nascent stage which is yet to see wider adoption and application in cybersecurity.

The field of malware detection has barely seen the presence of AutoML, and to the best of our knowledge has only been presented in a few works. Research in
\cite{kundu2021empirical} tested both AutoGluon-Tabular\footnote{https://auto.gluon.ai/stable/index.html} and Microsoft NNI\footnote{https://github.com/Microsoft/nni} on the EMBER-2018 dataset \cite{2018arXiv180404637A}, a malware dataset based on static analysis. These frameworks are used to tune hyper-parameters of a LightGBM model to best classify binaries from the dataset. Authors also used a proprietary dataset to evaluate the AutoML frameworks. 
This approach yielded a 3.2\% increase of True Positive Rate above the EMBER-2018 baseline results with the same classifier. AutoGluon-Tabular produced these results vs a 2.2\% increase with Microsoft NNI. Their approach largely used traditional machine learning methods as well as a FFNN in the ensemble offered with AutoGluon-Tabular. 
The authors of \cite{9658106_automl_enc_traffic} use AutoML to detect malware from encrypted network traffic. They used TLS fields as parameters to form their AutoML process. This work used a python package \textit{mljar-supervised}\footnote{https://supervised.mljar.com/}, utilizing many traditional ML models as well as a deep neural network in an ensemble.

%% file: sections/Static_AutoML.tex
\section{Automated Machine Learning for Static Malware Detection}\label{sec:static_automl}


\subsection{Deep Feed Forward Neural Networks}

Deep Feed Forward Neural Networks (FFNNs) are an extension of the simple perceptron network, except they often contain one or more hidden layers. FFNNs without convolutional or recurrent layers can also be referred to as Multi-Layer Perceptrons (MLPs).

\begin{figure}[!t]
\centering
\includegraphics[width=0.45\textwidth,keepaspectratio]{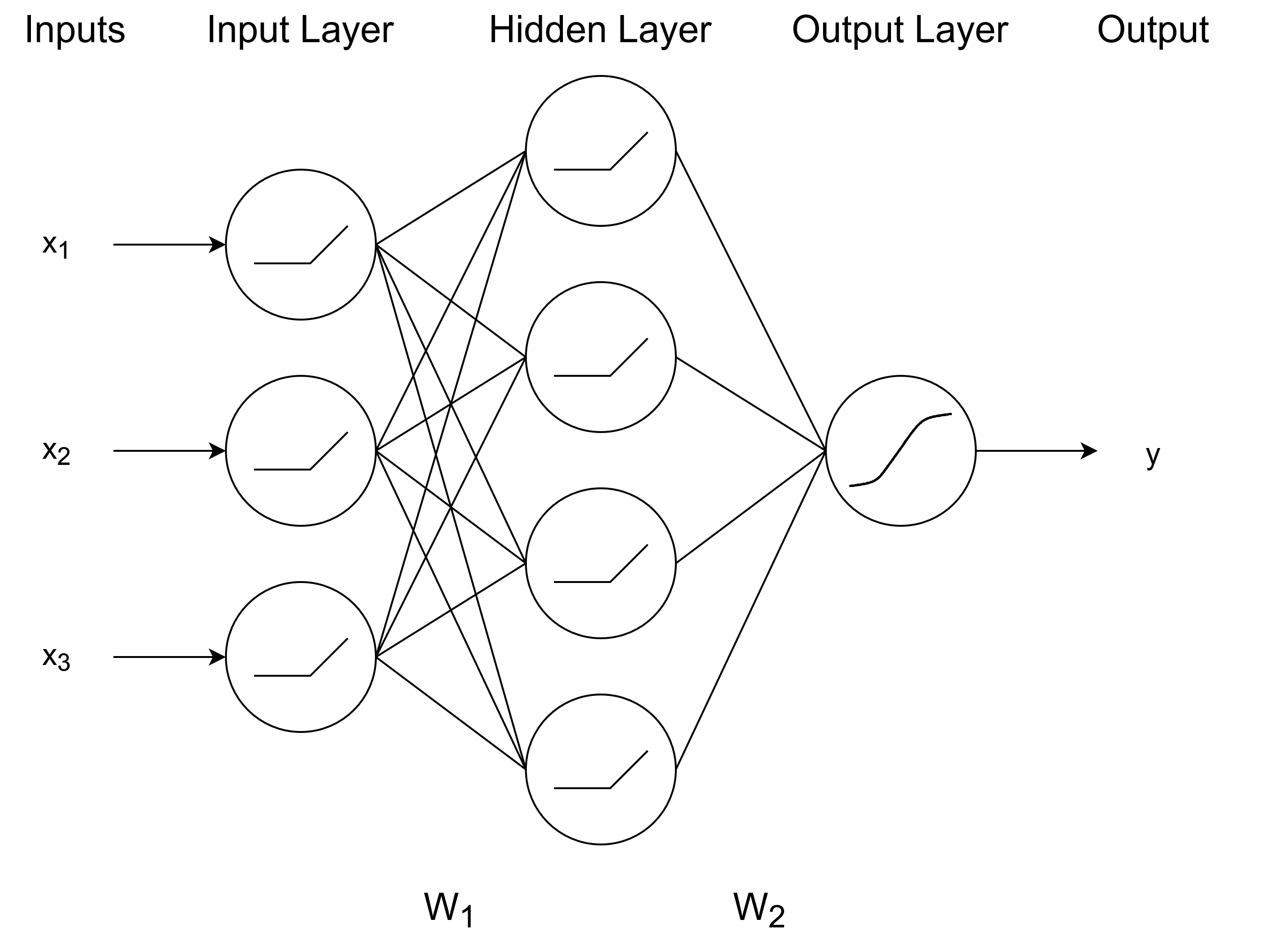}    
\caption{Feed Forward Neural Network}
\label{fig:ffnn}
\end{figure}

FFNNs pass input data through each layer in the model sequentially, applying each layer's function to the previous layer's output, forming what can be seen as an acyclic graph from input to output with data flowing only one way. Figure \ref{fig:ffnn} shows an example FFNN. Each node within a layer can apply an activation function to the sum of each of its inputs, shown as the function lines within each node in the figure. Each connection between nodes has a specific weight applied to the output of a specific node into another node. \(W_{1}\) and \(W_{2}\) represent the set of weights between each layer, each weight in each set corresponding to a connection between two unique nodes. The network can have any number of hidden layers. 

\begin{figure*}[!t]
\centering
\includegraphics[width=.8\textwidth]{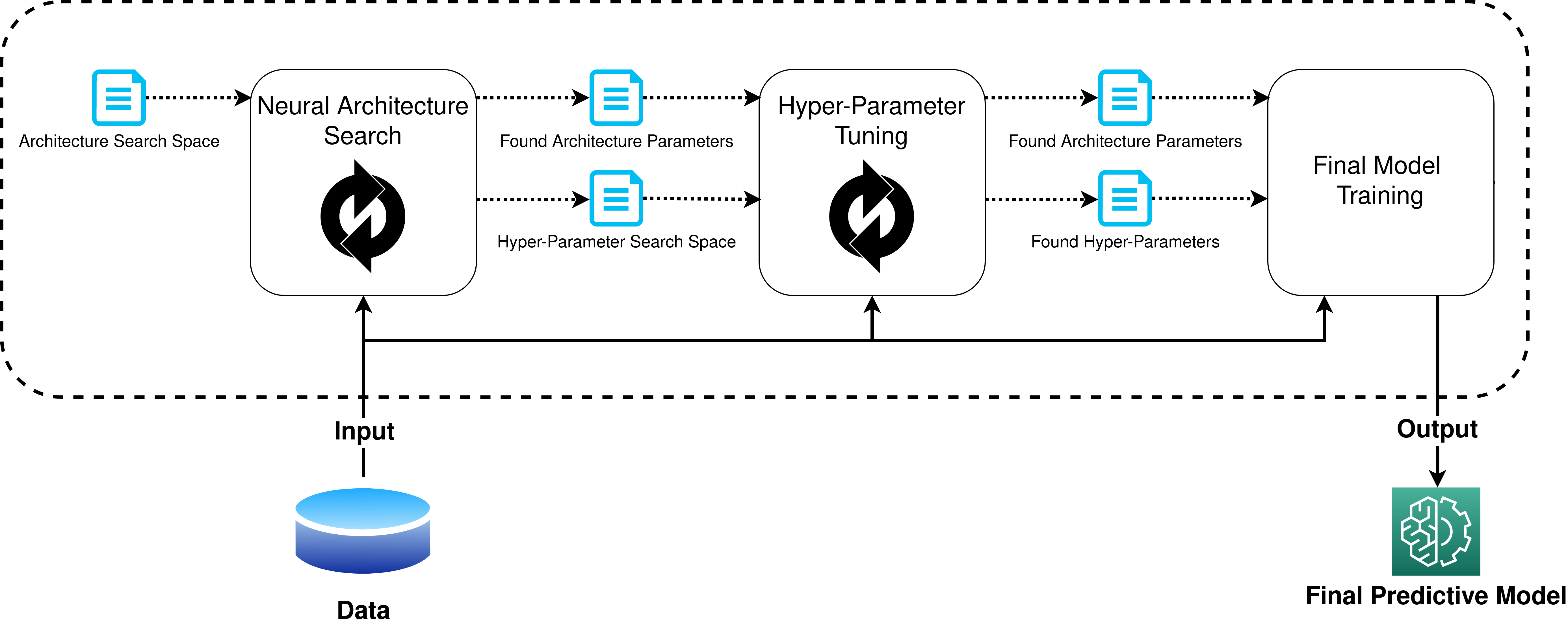}    
\caption{Automated Machine Learning Process}
\label{automl process}
\end{figure*}
Deep FFNNs require backpropagation through gradient descent to train weights of each layer of the network sequentially, backward through the network, from output to input. Through the processes of backpropagation, activation functions such as sigmoid and tanh can lead to a problem called vanishing gradients. This occurs because repeatedly taking the gradient of these functions, as backpropagation occurs, results in a value that approaches zero. For this reason, idempotent activation functions such as Rectified Linear Unit (ReLU) are often used in hidden layers of deep networks to solve the vanishing gradient problem. Additionally, functions like ReLU and Exponential Linear Unit (ELU) are cheaper to compute than sigmoid and tanh, but still allow for the network to learn non-linear functions. However, sigmoid like functions allow for an output to be mapped to a probability, and are often used on the output layer of a network to allow for each output neuron to produce a binary decision. Figure \ref{fig:ffnn} shows the input and hidden layer activation functions as ReLU, and the output layer's activation function as sigmoid.

\subsection{Search Methodology}
During the neural architecture search, an architecture selected from the search space is evaluated using an evaluation metric (F1-score our case), indicating the model performance based on which a strategy is employed to select the next architecture choice for subsequent evaluation. The next architecture selection in this work is based on a random selection strategy, where, regardless of the previous result, each new architecture choice is randomly selected without duplication. The NAS selects a number of random architecture configurations from the search space. These are referred to as trials. In each trial, a model is trained based on the selected architecture for a predefined number of epochs. Afterward, the model is evaluated at the end of every epoch using the evaluation score of the validation set. The model configuration that achieves the highest score will pass to the next phase, that is hyper-parameter tuning.

\subsubsection{Hyper-Parameter Tuning}
Once an architecture is selected during the NAS phase, the next phase, as shown in  Figure \ref{automl process}, searches for the optimal hyper-parameters of the chosen architecture. The hyper-parameters of the model are tune-able values that affect the model performance but do not alter the architecture of the model itself. This can include the batch size, optimizer, learning rate, dropout rate, etc. Just as in the NAS phase, the hyper-parameter tuning phase also has a bounded search space with a defined sampling granularity. With the hyper-parameter search space, we also define the sampling distribution.
The hyper-parameter search phase uses the Tree-structured Parzen Estimator (TPE) strategy \cite{bergstra2011algorithms} in selecting the next set of hyper-parameters to test.   

\subsubsection{Final Model Selection}
After the hyper-parameter tuning phase is complete, the results from the NAS phase and hyper-parameter tuning phase are combined to be the final model configuration. The model is then trained and evaluated after every epoch using the evaluation score of the validation set, and the highest performing epoch is saved as the final trained model to be evaluated on the test set.

\subsection{Static Malware Data Sources}
We use two popular static malware datasets - EMBER-2018 \cite{2018arXiv180404637A} and SOREL-20M \cite{harang2020sorel}, extensively used in the literature. We use these datasets with more primitive AutoML to explore the results of this methodology on datasets that have had high result benchmarks set.

\subsubsection{EMBER-2018 Dataset \cite{2018arXiv180404637A}}
EMBER is considered to be the first attempt to create an appropriately large static malware dataset. The dataset contains features extracted from benign and malicious portable executable (PE) files using the `Library to Instrument Executable Formats' (LIEF) \cite{LIEF}. The samples in the dataset are labeled as either malicious, benign, or unknown. Only the samples labeled malicious or benign are considered in this work. There are approximately 600K samples in the training set and 200K samples in the test set. Since there is no validation set provided, we excluded and used the last 20\% of the training set (i.e. 120K samples) as the validation set.
There are two versions of the EMBER dataset: EMBER-2017 and EMBER-2018. 
EMBER-2018 was specifically curated so that the training and testing sets would be harder to classify. We used EMBER-2018 in our experiments. However, to fairly compare our results to other works that used EMBER-2017, we test and report our model's (found with EMBER-2018) performance against EMBER-2017 dataset.

\subsubsection{SOREL-20M Dataset \cite{harang2020sorel}}
Sophos Labs\footnote{https://www.sophos.com/en-us/labs} released SOREL-20M dataset in 2020 to address some shortcomings of EMBER dataset. This dataset contains 12,699,013 training samples, 2,495,822 validation samples, and 4,195,042 testing samples. SOREL-20M uses the same features from the EMBER-2018 dataset.
The samples in the SOREL-20M dataset contain the same binary malicious label as EMBER-2018, but also contain extra metadata, including the number of anti-virus vendors that flagged a sample as malicious and the tags that anti-virus vendors associated with a sample. Included in these tags are labels such as \textit{dropper}, \textit{adware}, \textit{downloader}, etc. Authors in \cite{rudd2019aloha} have shown that the use of this metadata can help to improve performance, and our work in this section will allow the possibility of a model to use this auxiliary information in the training process.

\subsection{AutoML Tuning and Training}
\subsubsection{NAS Phase Configuration}
The full architecture search space for both the EMBER-2018 and SOREL-20M experiments is shown in Table \ref{architecture search space table}. The available options for \textit{Activation} and \textit{Tag Head Activation} are not applicable since the choices are either Rectified Linear Unit (ReLU) or Exponential Linear Unit (ELU). Similarly, for \textit{Use Counts} and \textit{Use Tags}, the choices are either \textit{True} or \textit{False}.

As mentioned previously, the SOREL-20M dataset has readily available labels each containing a binary malicious label, an encoding of the vendor tags, and a numerical count of the vendors flagging the sample as malicious. These additional labels were made available during the architecture search process through the use of additional output heads of the model to predict the count of the vendors flagging the sample malicious and predict any tags associated with the sample from anti-virus vendors. These additional heads were made optional through the use of two additional architecture search parameters: \textit{Use Counts} and \textit{Use Tags}, as shown in Table \ref{architecture search space table}. Design for additional heads, their respective loss functions, and the network design is inspired by \cite{rudd2019aloha}. 
The architecture search selects 150 random architecture configurations from the search space. The number of trials was chosen to cover both the search space and minimize cost. However, further investigation is required to analyze the effects of the number of trials on the selected models' performance as explained in subsection \ref{sub_sec:discussion}. The SOREL-20 and EMBER-2018 NAS was run for 10 and 25 epochs, respectively.

The highest achieved F1-score of a model during any point of its trial (instead of the F1-score of the final epoch) is chosen as the fitness score so that a model configuration's ability is more accurately represented, as the model's performance may fluctuate during the training process. Even though random search has been shown to give adequate results with a sufficient amount of trials \cite{random_search_high_dim, random_search_hyper_param}, trial count remains a parameter to be investigated in future work. 
\begin{table}[!t] \caption{Architecture Search Space}
\resizebox{0.5\textwidth}{!}
{
    \begin{tabular}{|c|c|c|c|}
    \hline
    \textbf{Parameter} & \textbf{\textit{Minimum}}& \textbf{\textit{Maximum}}& \textbf{\textit{Granularity}} \\
    \hline
    Depth& 1 & 14 & 1 \\
    \hline
    Width & 128 & 1920 & 128 \\
    \hline
    Activation & - & - & - \\
    \hline
    Tag Head Depth$^{\mathrm{*}}$ & 1 & 3 & 1\\
    \hline
    Tag Head Width$^{\mathrm{*}}$ & 16 & 112 & 16 \\
    \hline
    Tag Head Activation$^{\mathrm{*}}$ & - & - & - \\
    \hline
    Use Counts$^{\mathrm{*}}$ & - & - & - \\
    \hline
    Use Tags$^{\mathrm{*}}$ & - & - & - \\
    \hline
    \multicolumn{4}{l}{$^{\mathrm{*}}$SOREL-20M Models Only}
    \end{tabular}
}
\label{architecture search space table}
\end{table}

\subsubsection{Hyper-Parameter Tuning Phase Configuration}

\begin{table}[!t] \caption{Hyper Parameter Search Space}
\resizebox{0.5\textwidth}{!}
{
    \begin{tabular}{|c|c|c|c|c|}
    \hline
    \textbf{Parameter} & \textbf{\textit{Minimum}}& \textbf{\textit{Maximum}}& \textbf{\textit{Granularity}} & \textbf{\textit{Distribution}} \\
    \hline
    Batch Size (SOREL-20M) & 128 & 16384 & 1024 & quniform \\
    \hline
    Batch Size (EMBER) & 32 & 8192 & 32 & quniform \\
    \hline
    Learning Rate & 0.0001 & 1.0 & - & loguniform \\
    \hline
    Dropout & 0.0 & 0.50 & 0.05 & quniform\\
    \hline
    Tag Loss Weight$^{\mathrm{*}}$ & 0.0 & 1.0 & .05 & quniform \\
    \hline
    \multicolumn{4}{l}{$^{\mathrm{*}}$SOREL-20M Models Only}
    \end{tabular}
}
\label{hyper-paramater search space table}
\end{table}

The full hyper-parameter search space is shown in Table \ref{hyper-paramater search space table}.
 The \textit{quniform} distribution behaves like the sampling granularity in the NAS phase. The \textit{loguniform} samples from a logarithmic distribution such that the logarithm of the values returned will be uniformly distributed. Learning rate is sampled from this distribution to allow smaller values to be as likely sampled as larger values. We set the batch size minimum, maximum, and sampling granularity larger for the SOREL-20M experiments due to the size of the dataset as compared to EMBER-2018. Similar to the NAS phase parameters configuration, we believe that the minimum, maximum, and sampling granularity values requires further investigation.

 In \cite{rudd2019aloha}, using the SOREL-20M dataset, the authors use a loss weight of 0.1 for the vendor count head and vendor tag head, and a 1.0 loss weight for the malicious decision head. These loss weights can be considered a hyper-parameter available for tuning since altering the value does not change the architecture of the model. In our work, the malicious decision head loss weight is fixed to 1.0 while the auxiliary loss head weights are variable between 0.0 and 1.0 each. Note, only the tag head loss weight is included in Table \ref{hyper-paramater search space table} because the highest achieving model during the SOREL-20M NAS phase did not have a vendor count head, and therefore did not utilize that parameter.
 The models are trained for 10 and 25 epochs in the case of SOREL-20M and EMBER, respectively. F1-score is again used as the evaluation metric in selecting the highest performing model.

\subsection{Experimental Results}

\subsubsection{Evaluation Metrics}
We use four evaluation metrics along with receiver operating characteristic (ROC) and area under the curve (AUC).
\begin{equation}
Accuracy = \frac{TP + TN}{TP + TN + FP + FN}
\end{equation}
\begin{equation}
Precision = \frac{TP}{TP + FP} 
\end{equation}
\begin{equation}
Recall = \frac{TP}{TP + FN}  
\end{equation}
\begin{equation}
F1-score = 2 \times \frac{Precision \times Recall}{Precision + Recall}  
\end{equation}

Positive refers to a malicious sample, whereas, negative refers to a benign sample. \textit{TP}, \textit{FP}, \textit{TN} and \textit{FN} are true positives, false positives, true negatives and false negatives, respectively. Precision suffers when benign samples are labeled as malicious (high FP), while recall suffers when malicious samples are labeled as benign (high FN). F1-score is the harmonic mean of precision and recall, so it signifies models that have both high precision and recall. If a model has high precision but low recall or vice versa, the F1-score will be low. 

\subsubsection{Results}
After the experiments, using F1-score as an evaluation metric at each phase of the process, the found architectures and hyper-parameters are shown in Table \ref{found parameters}. 

\begin{table}[!t] \caption{Found Optimal Parameters}
\centering
\begin{tabular}{|c|c|c|}
\hline
\textbf{Parameter} & \textbf{SOREL-20M} & \textbf{EMBER-2018}\\
\hline
Depth & 8 & 3\\
\hline
Width & 1920 & 1664\\
\hline
Activation & ReLU & ReLU\\
\hline
Dropout & 0.15 & 0.30\\
\hline
Learning Rate & 0.000439 & 0.000269\\
\hline
Batch Size & 3072 & 1440\\
\hline
Use Count Head & False & - \\
\hline
Use Tag Head & True & - \\
\hline
Tag Head Depth & 1 & - \\
\hline
Tag Head Width & 112 & - \\
\hline
Tag Head Activation & ELU & - \\
\hline
Tag Head Loss Weight & 0.70 & - \\
\hline
\end{tabular}
\label{found parameters}
\end{table}

\begin{table*}[t] \caption{SOREL-20M Dataset Results}
\resizebox{\textwidth}{!}
{

\begin{tabular}{|c|c|c|c|c|c|c|c|}
\hline
\textbf{Work} &  \textbf{\textit{Perf. Metric}} & \textbf{\textit{AUC} }& \textbf{\textit{AUC $\leq$ 0.1\% FPR}} & \textbf{\textit{Accuracy}} & \textbf{\textit{F1-Score}} & \textbf{\textit{TPR: 0.1\% FPR}} & \textbf{\textit{TPR: 1\% FPR}} \\
\hline
\hline

ALOHA\cite{rudd2019aloha} & - & 0.997 & - & - & - & 0.922 & 0.972 \\
\hline

FFNN Ensemble\cite{extreme_false_positive} & - & \textbf{0.998} & 0.0927 & 0.988 & - & - & -\\
\hline

LightGBM Ensemble\cite{extreme_false_positive} & - & 0.984 & 0.0446 & 0.861 & - & - & - \\
\hline

Our Work & F1-Score & \textbf{0.998} & 0.966 & \textbf{0.990} & \textbf{0.984} & \textbf{0.965} & 0.993\\
\hline

Our Work & Loss & \textbf{0.998} & \textbf{0.969} & \textbf{0.990} & \textbf{0.984} & 0.963 & \textbf{0.995}\\
\hline

\end{tabular}
\vspace{-4mm}

}
\label{SOREL-20M results table}
\end{table*}

\begin{table*}[t] \caption{EMBER Dataset Results}
\resizebox{\textwidth}{!}
{

\begin{tabular}{|c|c|c|c|c|c|c|c|}
\hline
\multicolumn{8}{|c|}{\centering\textbf{EMBER 2018}}\\

\textbf{Work} &  \textbf{\textit{Perf. Metric}} & \textbf{\textit{AUC} }& \textbf{\textit{AUC $\leq$ 0.1\% FPR}} & \textbf{\textit{Accuracy}} & \textbf{\textit{F1-Score}} & \textbf{\textit{TPR: 0.1\% FPR}} & \textbf{\textit{TPR: 1\% FPR}} \\
\hline
\hline

AutoGluon Ensemble \cite{kundu2021empirical} & - & - & - & - & - & \textbf{0.900} & - \\
\hline

Malconv w/ GCG\cite{raff2020classifying} & - & 0.980 & - & 0.933 & - & - & - \\
\hline

LightGBM Ensemble\cite{extreme_false_positive} & - & 0.986 & 0.0605 & 0.940 & - & - & -\\
\hline

Detection Pipeline\cite{loi2021towards} & - & \textbf{0.995} & - & \textbf{0.969} & - & - & - \\
\hline

Our Work & F1-Score & 0.984 & \textbf{0.614} & 0.958 & \textbf{0.958} & 0.417 & \textbf{0.969}\\
\hline

Our Work & Loss & 0.981 & 0.573 & 0.918 & 0.921 & 0.188 & 0.951\\
\hline

\hline
\hline
\multicolumn{8}{|c|}{\centering\textbf{EMBER 2017}}\\
\textbf{Work} &  \textbf{\textit{Perf. Metric}} & \textbf{\textit{AUC} }& \textbf{\textit{AUC $\leq$ 0.1\% FPR}} & \textbf{\textit{Accuracy}} & \textbf{\textit{F1-Score}} & \textbf{\textit{TPR: 0.1\% FPR}} & \textbf{\textit{TPR: 1\% FPR}} \\
\hline
\hline

DeepMalNet\cite{deepmalnet_journal} & - & - & - & 0.989 & 0.989 & - & - \\
\hline

MalConv\cite{malconv} & - & - & - &  0.988 & 0.988 & - & - \\
\hline

Our Work & F1-Score & \textbf{0.999} & \textbf{0.916} & \textbf{0.992} & \textbf{0.992 }& \textbf{0.956} & \textbf{0.997}\\ 
\hline

\end{tabular}

}
\label{EMBER results table}
\end{table*}

The detection results are listed in Table \ref{SOREL-20M results table} and Table \ref{EMBER results table} for SOREL-20M and EMBER datasets, respectively. Also included in this table is the AUC with a maximum false positive rate (FPR) of 0.1\%, the accuracy, F1-score, true positive rate (TPR) at 0.1\% FPR, and TPR at 1\% FPR. The table also contains results using loss as a performance metric for SOREL-20M and EMBER-2018; this is to show the difference in F1-score and loss as a performance metric in the final stage, this will be brought up in the discussion section of this section. Some other works shown in the tables only report a subset of the metrics, but are still shown for comparison. 

In particular, for SOREL-20M, Table \ref{SOREL-20M results table} shows our AUC results are on par with the FFNN ensemble from \cite{extreme_false_positive} and slightly exceeds \cite{rudd2019aloha}, the work that presented the auxiliary model heads for SOREL-20M.  Our model significantly exceeds
the AUC under 0.1\% FPR of the only other work \cite{extreme_false_positive}, which reported this parameter. The accuracy of our model is similar but higher than \cite{extreme_false_positive}. We reported TPR at 0.1\% and 1\% FPR for comparison to \cite{rudd2019aloha}, where it can be seen our model performed better in both. 

With respect to EMBER-2018 in Table \ref{EMBER results table}, \cite{loi2021towards} performs slightly better in their reported metrics, AUC and accuracy, whereas the rest of their metrics are not reported. 
Our model chosen in the final training phase is similar to other results in AUC and accuracy, surpassing \cite{raff2020classifying} in AUC, and surpassing both \cite{raff2020classifying, extreme_false_positive} in accuracy. The AUC under 0.1\% FPR of our model far surpasses the results of \cite{extreme_false_positive}.  The TPR at 0.1\% is the only reported metric of \cite{kundu2021empirical}, which is significantly higher than our results. Due to limited metrics provided by other related works, it is difficult to compare the efficacy of our AutoML method in a holistic sense. 
The results from EMBER-2017 (with the optimal parameters from EMBER-2018) are reported in the bottom of Table \ref{EMBER results table}. The authors in \cite{deepmalnet_journal} and \cite{malconv} only reported accuracy and F1-score of their results. Our model's accuracy and F1-score are slightly higher than their results, but with metrics close to 100\%, this is significant.

The results show that models developed with our proposed AutoML pipeline are similar to those found with hand designed solutions, and sometimes even exceed the performance. This shows the efficacy of integrating AutoML into a malware detection pipeline, eliminating the need to  hand designed models, which is difficult, time consuming, and requires high technical skills.

\begin{figure}[!t]

\begin{center}
\includegraphics[width=0.5\textwidth]{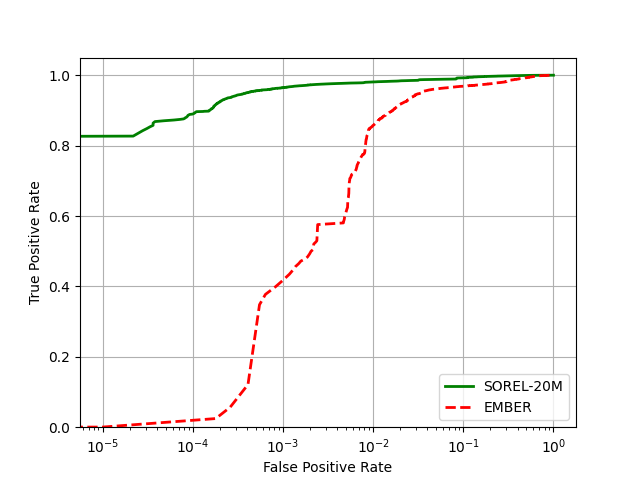}
\end{center}
\caption{ROC using F1-Score for Selection}
\label{roc f1}
\end{figure}

Figure \ref{roc f1} shows the ROC curve for both the EMBER-2018 and SOREL-20M experiments, note the logarithmic scale on the x-axis denoting the FPR. An ROC curve shows the TPR for each respective FPR. Given the magnitude increase of training data in SOREL-20M over EMBER-2018, it is no surprise the TPR of SOREL-20M is higher than EMBER-2018 at any FPR. SOREL-20M TPR falls off much slower than EMBER-2018, and never goes below 0.8 TPR in the graph.

The training and evaluation were performed on a virtual machine equipped with 92 vCPUs and 448 GB of memory. Additionally, it incorporated 8 Tesla V100 GPUs, each with 16 GB of VRAM. While the number of cores and memory might have been excessive for the task at hand, the presence of 8 GPUs facilitated parallel training sessions for neural network hyperparameter optimization.

\subsection{Discussion and Analysis}
\label{sub_sec:discussion}

\subsubsection{Meta-Hyper-Parameter Selection}
As mentioned earlier, many of the parameters governing the NAS and Hyper-Parameter tuning phases are selected based on our experience to simplify the process and provide a balance between the cost of training and detection results. We discuss below some of these parameters.  

\subsubsection{Epochs per Trial}
The number of epochs per trial is an important parameter that directly affect the NAS process.
This was especially a consideration for the SOREL-2OM trials, since the dataset is an order of magnitude larger than the EMBER-2018 dataset and therefore took much longer to train. 

\begin{figure}[!t]

\begin{center}
\includegraphics[width=0.45\textwidth]{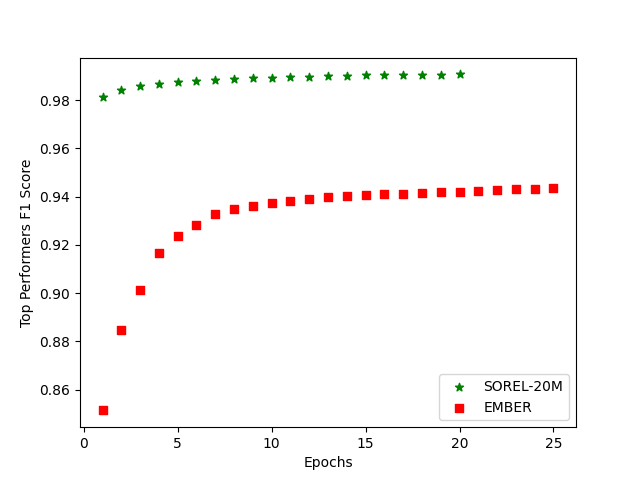}
\end{center}
\caption{Top 30 Preforming Models Average F1 by Epoch}
\label{epoch_v_f1}
\end{figure}

Initially, the number of epochs for the SOREL-20M NAS trials was set to 3. This implies that the model configurations with the highest performance after training for 3 epochs would perform the best overall. To test this, we increased the number of epochs to 10 and 20 to help in better understanding of the impact of epochs per trial on the selected models' performance during the NAS. The results of these experiments are shown in Figure \ref{epoch_v_f1}. The primary consideration here is with the performance trend of SOREL-20M, but EMBER-2018 is shown as well.

\begin{figure}[!t]

\begin{center}
\includegraphics[width=0.45\textwidth]{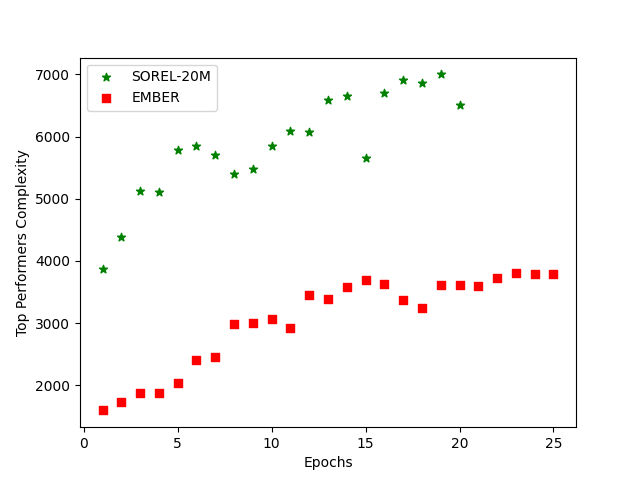}
\end{center}
\caption{Top 30 Preforming Models Average Complexity by Epoch}
\label{epoch_v_complexity}
\end{figure}

This graph shows the F1-score average of the top 30 selected models at any given epoch. The F1-score for each model is calculated as the highest F1-score reached up to and including a given epoch. At each epoch, the 30 models with the highest F1-score, as described above, are averaged together. At any epoch, the top 30 set of models may be different if any model in the experiment achieves results that puts the model in the top 30 for that epoch. 
It can be seen that there is a correlation between top model performance and the number of epochs, in a seemingly logarithmic relationship. As long as a model is not so complex that it over-fits the training data, a more complex model should, intuitively, preform as well as or better than a less complex model. However, a good choice for the number of epochs should be where the curve start to straighten so that the model doesn't become too complex and, in turn, require a massive amount of training time.
Figure \ref{epoch_v_complexity} shows that as the number of training epochs per trial increases so does the average complexity of the top 30 performing models.
The model complexity in the figure represents the average product of \textit{width} and \textit{depth} of the hidden layers of the top 30 model configurations during the NAS phase, which results in the number of trainable parameters in a given model.

\subsubsection{NAS and Tuning-parameters Phases Evaluation Metric}
\begin{figure}[!t]

\begin{center}
\includegraphics[width=0.45\textwidth]{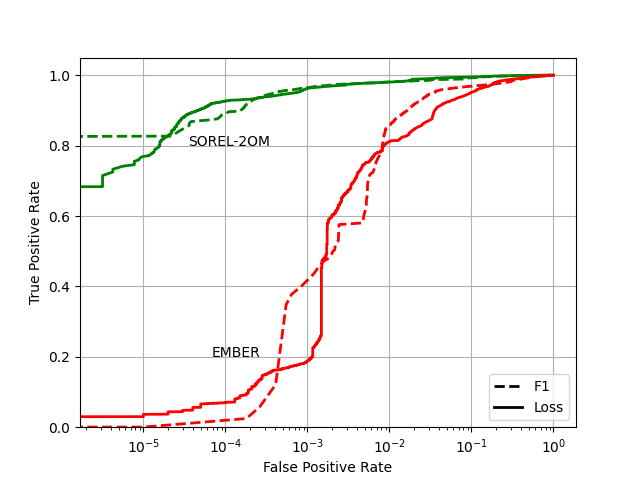}  
\end{center}
\caption{ROC using Loss vs F1 for Selection}
\label{roc_loss_and_f1}
\end{figure}

We use F1-score as an evaluation metric to select the models that have both high recall and precision. 
After getting the final selected model during the NAS and hyper-tunning-parameter phases, we train and evaluate the model using both F1-score and binary cross-entropy loss. The results shown in the ROC curves in Figure \ref{roc_loss_and_f1} shows that both metrics can reach comparable results. This indicates that, besides F1-score, other metrics could also be explored, including accuracy, AOC, binary cross-entropy loss, etc. and are left to future work. 


\subsubsection{Search Space Bounding and Strategy}
The search space values are one of the most important factors in the AutoML process. As shown in Table \ref{found parameters}, the width parameter (i.e. 1664) of selected model for EMBER-2018 dataset is found to be less than the maximum value (i.e. 1920). However, the width of the SOREL-20M model was the maximum available value in the bounded search space. This indicates that an even wider model might preform better than the found model had the search space been bigger. Selecting optimal search space values is still an open question. 
The choice of the search strategy for the NAS and tuning-parameter phases are random search and TPE, respectively. In this section, these strategies were chosen because of their simplicity. However, a more adequate strategy tailored to malware detection could potentially result in better selected AutoML models.


\subsubsection{Cost of Current Implementation}
The SOREL-20M experiments took $\approx$30 minutes per epoch to run, with 16 experiments running simultaneously. The EMBER-2018 experiments took $\approx$5 minutes per epoch, with 24 experiments running simultaneously. Overall, the experiments took $\approx$5600 minutes and $\approx$1560 minutes to run both SOREL-20M and EMBER-2018 experiments, respectively. This is the time to run the NAS and hyper-parameter phases of the process, excluding the final model training. The time to train the final model is not reported, as the computational cost is insignificant compared to the previous two phases.

The current implementation of the proposed methodology uses a \textit{multi-trial NAS}, where each set of model parameters selected from the NAS search space are trained to the specified epoch limit. Other implementations of multi-trial NAS try to optimize this process through early stopping and weight sharing \cite{li2020random}. Even though these methods may introduce instability into the process, they can reduce the computational cost. 

It can be concluded that it is more expensive to use AutoML than to train hand-designed models. This cost trade-off should be taken into account as the proposed methodology becomes more refined. Future implementations may significantly reduce the time to complete the AutoML process. This can be achieved through more sophisticated NAS implementations and intelligent search strategies that can reduce the number of trials required or the number of epochs required per trial.

%% file: sections/Online_AutoML.tex
\section{Automated Machine Learning for Online Malware Detection}\label{sec:online_automl}
\begin{figure*}[!t]
      \centering
    \includegraphics[keepaspectratio,width=0.7\textwidth]{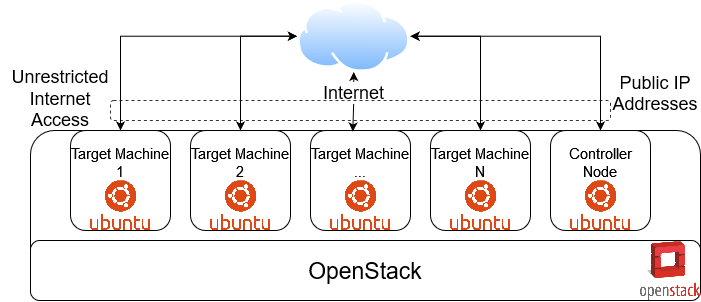}
     \caption{Cloud Testbed Setup}
    \label{testbed}
\end{figure*}
This section will focus on using one-shot AutoML for malware detection in online cloud environments using Convolutional Neural Networks (CNNs). 

\subsection{Convolutional Neural Networks}
CNNs are a widely used type of deep learning model designed for image type data. CNNs work differently than regular deep FFNNs, where the output of every node is passed into every node of the next layer. CNNs receive a 3 dimensional input (channels,height,width). CNNs have filters, whose values are learned, that convolve across input channels to detect edges. The primitive edges detected in earlier layers can be combined in later layers to learn more complex shapes. The core of CNN layers fall into two categories: normal and reduction convolutional layers. Normal convolutional layers use filters to convolve across the input to produce data with more channels, keeping the same height and width. Reduction cells to reduce the width and height of its input data to reduce the number of trainable parameters in the next layer or cell. The output of the convolutional layers is passed through a pooling layer to reduce the input dimension to 1 for dense layers to produce the network output (prediction). 

CNNs are used in this section because process performance metric data can be grouped together in the form of an image, with rows denoting unique processes and columns denoting performance features of these processes.


\begin{figure}[!t]
 \centering
\includegraphics[keepaspectratio,width=0.45\textwidth]{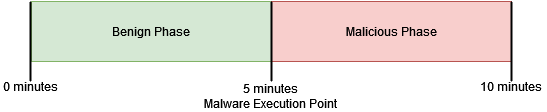}    
     \caption{Experiment Phases}
    \label{experiment}
\end{figure}

\subsection{Online Cloud Testbed}
 Figure \ref{testbed} illustrate the testbed utilized to generate the online malware dataset in an OpenStack\footnote{https://www.openstack.org/} instance hosted by the University of Texas at San Antonio.  All virtual machines used to create this dataset had open and unrestricted internet access, as well as a public IP address.  Each virtual machine is running a fully up-to-date Ubuntu 18.04 instance.  The experiments are controlled and data gathered by a controller node within the OpenStack testbed.  Each VM contains programs to collect data from their respective sources, which at the end of the experiment is collected by the controller node.  Before each experiment, each target VM is reset to a clean state.  Each virtual machine has 2 CPU cores, 4 gigabytes of RAM, and 40 gigabytes of disk space.  

\begin{figure}[!t]
        \begin{center}
\includegraphics[keepaspectratio,width=0.45\textwidth]{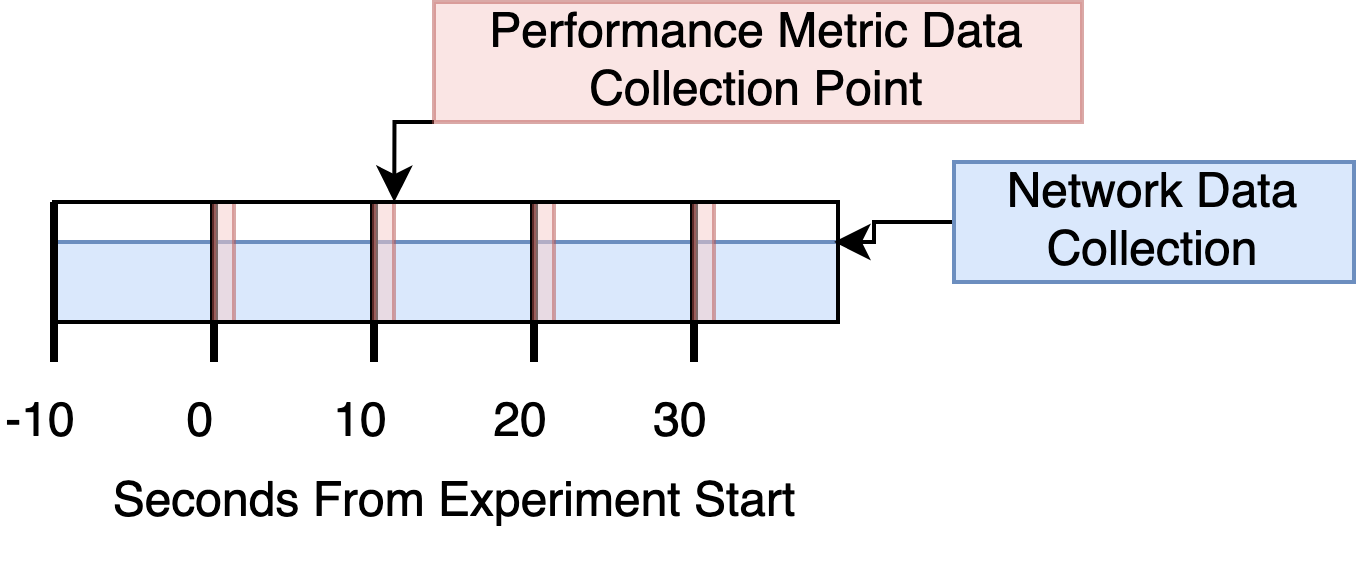}  
    \end{center}
    \caption{Data Collection Phases}
    \label{collection}
\end{figure}
\subsection{Application and Baseline Sets}
To best understand the behavior of malware on a full, online system, it may help to include malware data when the machine is idle and fully operational. For the purposes of this dataset, the fully operational server will be an Apache web server hosting a WordPress application, with a MySQL database on the backend. To model real world end users of the server, an on ON/OFF Pareto distribution following NS2\footnote{http://www.isi.edu/nsnam/ns/doc/node509.html} parameters is utilized to mimic the distribution of client requests to the webserver. All malware was run with only user level privileges.

\subsubsection{Malware Source and Selection}
The malware selected for this data came from a variety of sources, including VirusTotal\footnote{https://www.virustotal.com/}, MalShare\footnote{https://www.malshare.com/}, VirusShare\footnote{https://virusshare.com/}, Linux-Malware-Samples\footnote{https://github.com/MalwareSamples/Linux-Malware-Samples}, and MalwareBazarr\footnote{https://bazaar.abuse.ch/}. The gathered samples were tested for ability to execute on the target hardware in case the mutable header field of the malware had been altered, in which case the malware may not run on the target hardware. Also, samples that lead to corruption of the collected data during the experimentation process were removed from consideration after the fact. In total 4077 malware samples were considered.

\subsubsection{Data Collection}
The experiment length for this dataset is 10 minutes - meaning data is collected for the entirety of 10 minutes.  Halfway through a given experiment, the malware being tested is executed.  Therefore, every experiment contains an equal amount of benign and malicious activity.  This can be seen in Figure \ref{experiment}. There are multiple random benign SSH connections made to each target box throughout the experiment to mask the SSH connection used to spawn the malware execution.

\begin{figure}[!t]
\begin{center}
\includegraphics[width=0.4\textwidth]{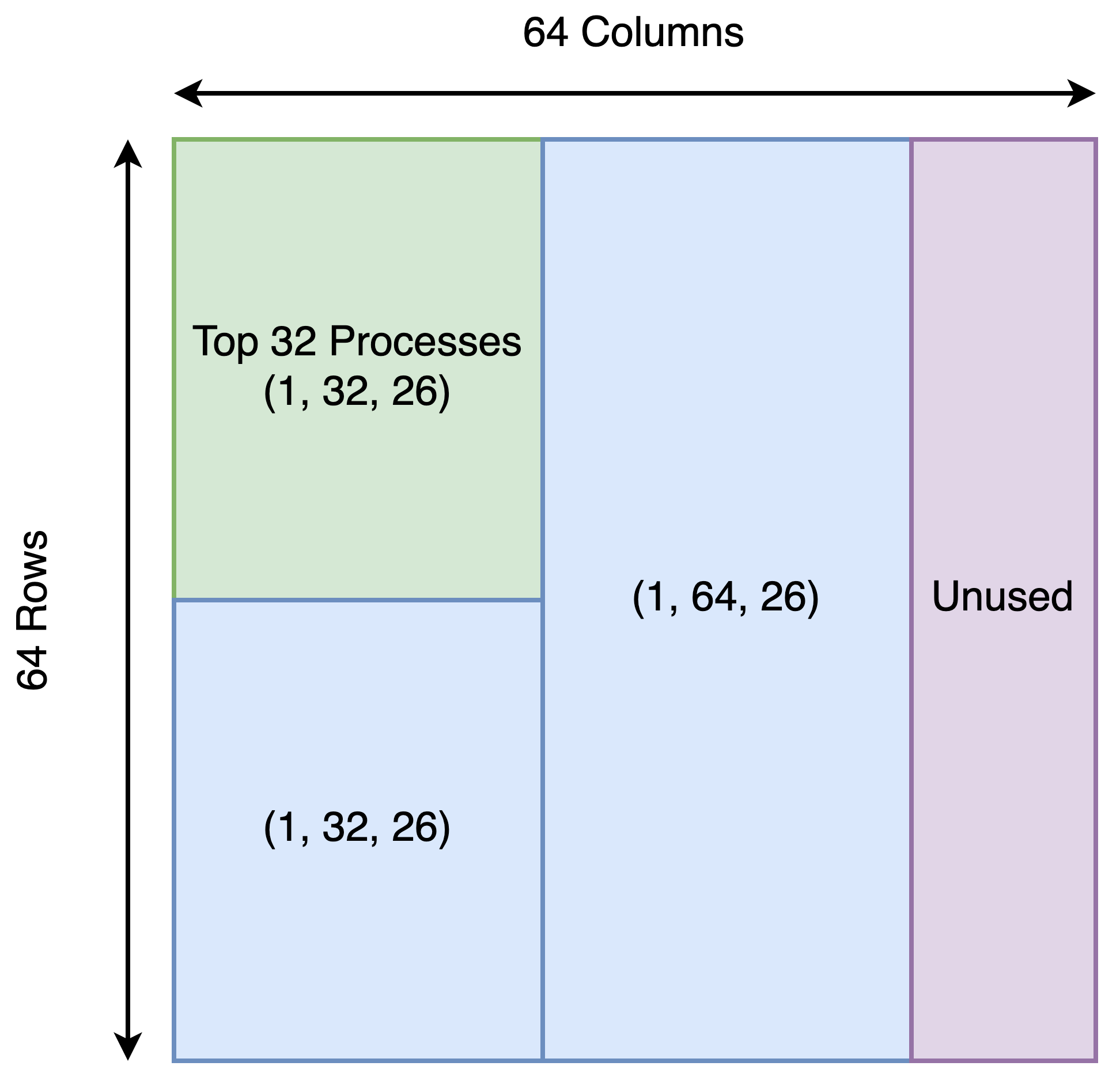}
\end{center}
\caption{Input Data Shape}
\label{data shape}
\end{figure}

The methods by which different sources of data are collected contain both continuous and discrete collection.  Network data is collected continuously throughout the experiment, and starts 10 seconds early to allow for a delta to be taken, since the collection is a running total of network activity per process.  Per-process data is collected at an interval of every 10 seconds, taking the instantaneous value of the monitored metrics.  The collection over time for each data source as shown in Figure \ref{collection}. Specifics of each type of data collected will be discussed in the following subsections.

\subsubsection{Per-Processes Performance Data}
Performance metrics are collected on a per-process basis. This data is collected every 10 seconds for the duration of the experiment.  The python library \verb|psutil| is used to collect this data.

Process IDs (PIDs) would, at first, seem like an easy way to identify a unique process thought the experiment, but this doesn't hold true.  A Linux kernel by default has a maximum PID of 32768, at which point PIDs begin getting re-used.  Therefore, it is feasible that in a highly active system that creates many new processes and closes old ones, that a single PID may identify more than one process during the experiment run-time.  Instead, a tuple of the entire command line (including arguments) of the process and a hash of the executable (if applicable) is collected.  This is much less likely to collide with the identifier of another process.  

\begin{figure}[!t]
\begin{center}
\includegraphics[height=0.35\textwidth, angle=90]{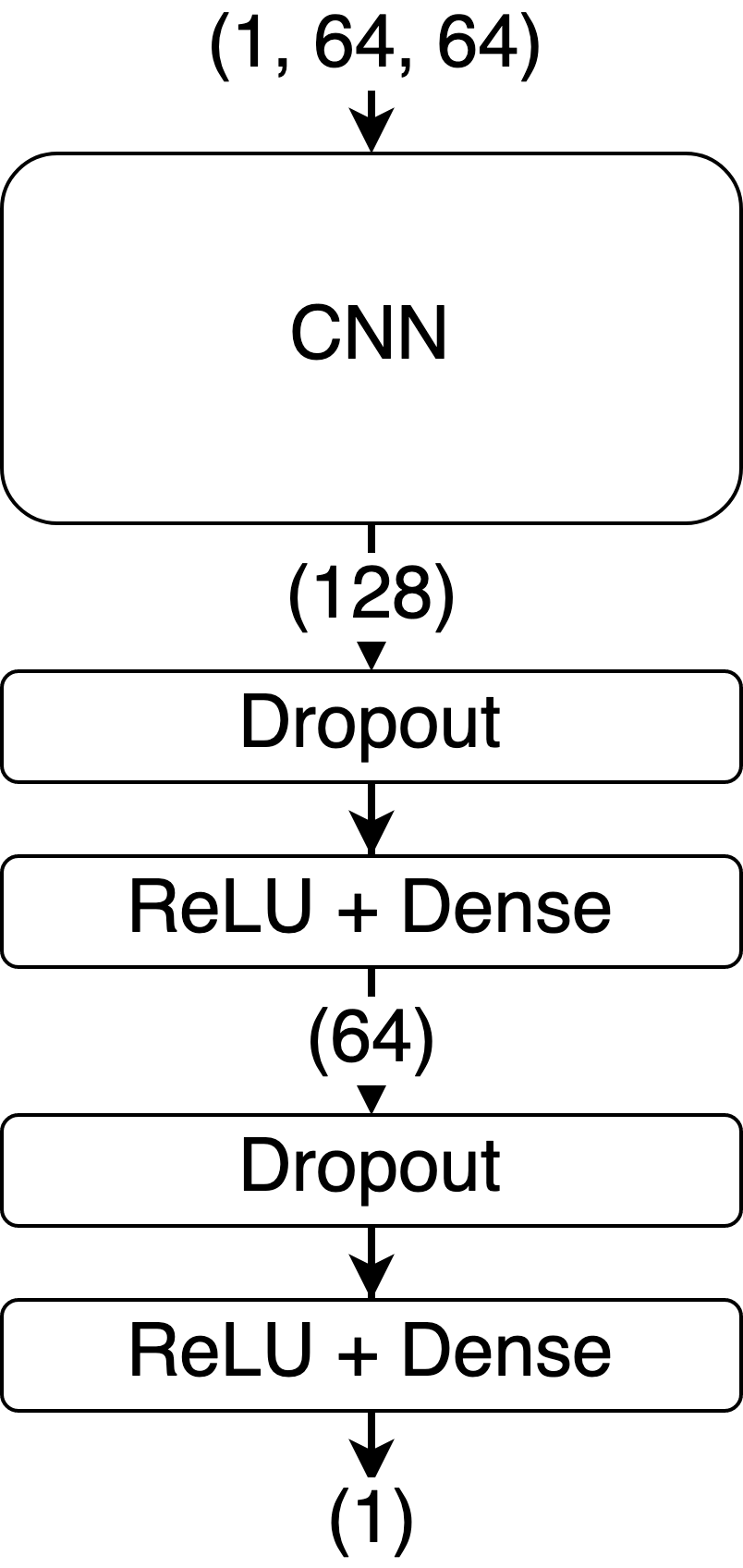}    
\end{center}
\caption{General Architecture}
\label{achitecture}
\end{figure}
\subsubsection{Per-Process Network Data}
Many data collection tools do not allow for the collection of network traffic statistics in a per-process basis.  However, the tool Nethogs\footnote{https://github.com/raboof/nethogs} allows for the grouping of bandwidth by process, and is used to collect network bandwidth data in the experiments.  A python wrapper is used to interact with the Nethogs library for data collection.

The network bandwidth data (bytes in/out) per process is recorded as a running total, therefore network data collection is started 10 seconds early, and the delta between each record is used in post-processing.  In order to match network data to process data, the PID at a given timestamp in the network data can be compared to the records in the process data, which ultimately holds the primary key to denote a unique process.

\subsubsection{Combined Data and Representation}
 In order to include network data with per-process performance metrics, the data is combined. First, any record of the data collection agents is removed from the per-process performance data.  The data that is left in process data will be the basis by which network usage is searched in network data. The discrete process data is grouped by collection time (every 10 seconds), and any matching network data between collection times is added to the latter process data collection record. That is for a given unique process record \textit{p} taken at collection time \textit{N}, any matching network data records for \textit{p} between the previous collection time \textit{N-1} and current collection time \textit{N}, will be added to the process record of \textit{p} at collection time \textit{N}. A sample feature table for a unique process is shown here in Table \ref{tab:perf_metrics}.

\begin{table*}[h]
\caption{Features Sample}
\resizebox{\textwidth}{!}
{
    \begin{tabular}{|c|c|c|c|c|c|}
        \hline
        \textbf{Metric} & \textbf{Value} & \textbf{Metric} & \textbf{Value} & \textbf{Metric} & \textbf{Value}\\
        \hline
        num\_fds & 78 & cpu\_percent & 0.0 & cpu\_time\_user & 0.15\\
        cpu\_time\_system & 1.7 & cpu\_time\_children\_user & 7.64 & cpu\_time\_children\_system & 3.1\\
        context\_switches\_voluntary & 1390 & context\_switches\_involuntary & 430 & num\_threads & 1\\
        memory\_info\_rss & 9113600 & memory\_info\_vms & 163598336 & memory\_info\_shared & 6795264\\
        memory\_info\_text & 1376256 & memory\_info\_lib & 0 & memory\_info\_data & 18956288\\
        memory\_info\_dirty & 0 & memory\_info\_pss & 2922496 & memory\_info\_swap & 0\\
        io\_read\_count & 53242& io\_write\_count & 18782 & io\_read\_bytes & 320275456\\
        io\_write\_bytes & 113713152& io\_read\_chars & 248760749 & io\_write\_chars & 152977520\\
        sent\_bytes & 0.0 & recv\_bytes & 0.0 & & \\
        
        \hline
    \end{tabular}
}
\label{tab:perf_metrics}
\end{table*}

To feed the data to models, the data is represented as a single channel (grayscale) image. The columns of this image are the collected performance metrics and the rows are unique processes. As shown in Figure~\ref{data shape}, the image dimensions are represented as (channels, rows, columns) and are selected to be (1, 64, 64). The first 26 columns and second 26 columns each contain performance metrics for the rows of processes. That is, the 56 used columns of the input are divided into two sections of 26 features, each of these meta-columns representing a unique processes features. The first 32 rows of the first meta-column are reserved for commonly occurring processes found in the training set, so that in every input sample, a process that is commonly occurring will be in the same spot in the data in every input sample.

There are 12 blank columns, padded with 0, on the right side of the image that are used as padding so the image can maintain a square shape. The image shape is selected to be square and a power of 2 to ensure there are no dimensionality problems when feeding the data into a variety of CNN models. A total of 128 unique processes can be included in an image, and the top 32 processes that occur very frequently throughout the data will always be placed in the same row and column throughout all samples.

\vspace{-2mm}

\subsection{Methodology}
We used one-shot learning to find a performant CNN 
to detect malware from the performance metric data. The Darts \cite{liu2018darts} AutoML methodology is applied to search for an optimal CNN architecture from the training data. The code for this is adapted from the Microsoft NNI implementation of Darts. Darts works to find normal and reduction convolutional cells by figuring out layer connections between nodes in the repeated cells. The found architecture will be a normal and reduction convolutional layer in a CNN with a specified number of layers (cells), nodes per cell, and channels per node. Increasing the number of nodes, and even more so increasing the number channels per node, can create large memory overhead in the neural architecture search. Darts finds the connections between nodes in a cell by posing the probability of a connection being the best as a softmax, so the best connections can be found using gradient descent. For further explanation on the DARTS AutoML process, refer to the original paper \cite{liu2018darts}.

The choices for connections between nodes in a cell are \textit{skip connect} (identity for normal cells and factorized reduction for reduction cells), \textit{dilated convolution} (5x5 or 3x3), \textit{separable convolution} (3x3 or 5x5), \textit{average pooling} (3x3), or \textit{max pooling} (3x3). These were the choices in the original Darts paper and are also used here. Stochastic Gradient Descent (SGD) optimizer and a learning rate scheduler are both used, with the same parameters as described in \cite{liu2018darts}. 


The general architecture for the entire network is shown in Figure  \ref{achitecture}. The CNN part of the model is either be found by with the Darts methodology or is a state-of-the art CNN for comparison. 
\begin{figure}[!t]
\begin{center}
\includegraphics[width=0.35\textwidth]{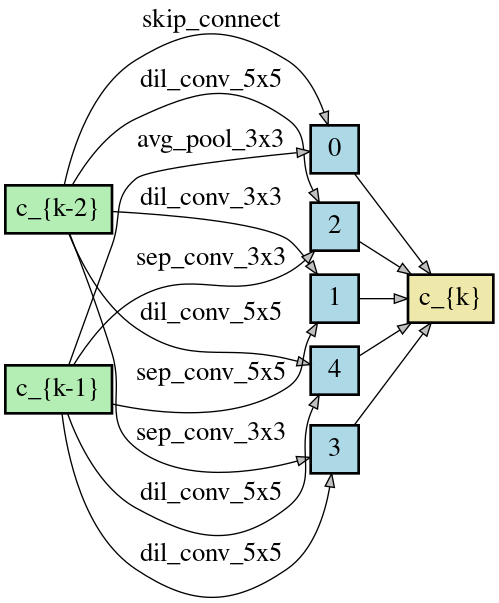}    
\end{center}
\caption{Found Normal Cell}
\label{normal cell}
\end{figure}
\subsection{Training and Results}

\subsubsection{Data Splits}
Given 4077 total malware experiments per set (baseline/application), running for 10 minutes each, with data points at every 10 seconds, 246,620 total samples are available in the baseline and application dataset. 80\% of the experiments are used for training, 10\% for validation, and 10\% for testing. No experiment (malware sample) is contained in more than one set (training/validation/testing). Also, the baseline training set consists of the same malware as the application training set, and the same is true for validation and test sets. A mean and standard deviation are calculated using the training set in the baseline and application set, and is used to normalize each of the respective datasets.

\begin{figure*}[!t]
\begin{center}
\includegraphics[width=0.8\textwidth]{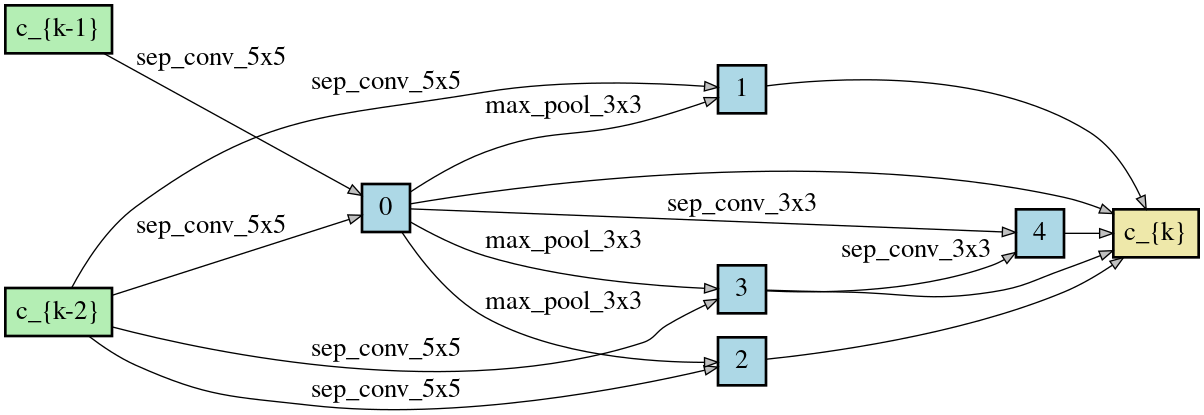}    
\end{center}
\caption{Found Reduction Cell}
\label{reduction cell}
\end{figure*}

\begin{table*}[t] \caption{Online Detection Results}
\resizebox{\textwidth}{!}
{

\begin{tabular}{|c|c|c|c|c|c||c|c|c|}
\hline
\multicolumn{9}{|c|}{\centering\textbf{Baseline}}\\

\textbf{Model} &  \textbf{\textit{Accuracy}} & \textbf{\textit{Precision}}& \textbf{\textit{Recall}} & \textbf{\textit{F1-Score}} & \textbf{\textit{AUC}} & \textbf{\textit{Delay @ Low FPR}} & \textbf{\textit{TPR @ Low FPR}} & \textbf{\textit{FPR @ Low FPR}}\\
\hline
\hline

Resnet18 & 0.97463 & \textbf{0.99387} & 0.95511 & 0.97411 & 0.99877 & 10.56373 s& 0.96321 & 0.00735 \\ 
\hline

Resnet50 & 0.97913 & 0.96266 & \textbf{0.99689} & 0.97948 & 0.99892 & 9.60784 s& 0.96681 & 0.00759 \\ 
\hline

Resnet101 & 0.98897 & 0.98856 & 0.98937 & 0.98897 & 0.99927 & 3.60294 s& 0.98814 & 0.01045 \\ 
\hline

Densenet121 & 0.98358 & 0.99079 & 0.97621 & 0.98344 & 0.99896 & 7.81863 s& 0.97367 & 0.00816 \\ 
\hline

Densenet169 & 0.98346 & 0.97972 & 0.98733 & 0.98351 & 0.99838 & 6.66667 s& 0.97490 & 0.01086 \\ 
\hline

Densenet201 & 0.98570 & 0.99148 & 0.97981 & 0.98561 & 0.99907 & 5.90686 s& 0.98005 & \textbf{0.00898} \\ 
\hline

Darts AutoML & \textbf{0.98917} & 0.98674 & 0.99166 & \textbf{0.98919} & \textbf{0.99954} & \textbf{3.03922 s} & \textbf{0.98986} & 0.01094 \\ 
\hline

\hline
\hline
\multicolumn{9}{|c|}{\centering\textbf{Application}}\\
\textbf{Model} &  \textbf{\textit{Accuracy}} & \textbf{\textit{Precision}}& \textbf{\textit{Recall}} & \textbf{\textit{F1-Score}} & \textbf{\textit{AUC}} & \textbf{\textit{Delay @ Low FPR}} & \textbf{\textit{TPR @ Low FPR}} & \textbf{\textit{FPR @ Low FPR}}\\
\hline
\hline

Resnet18 & 0.96246 & 0.94401 & 0.98709 & 0.96507 & 0.99417 & 21.33995 s& 0.92964 & 0.01945 \\ 
\hline

Resnet50 & 0.96667 & 0.96196 & 0.97512 & 0.96850 & 0.99480 & 14.54094 s& 0.94084 & 0.01541 \\ 
\hline

Resnet101 & 0.97953 & 0.97627 & 0.98500 & 0.98061 & 0.99728 & 10.19851 s& 0.96470 & 0.01446 \\ 
\hline

Densenet121 & 0.97239 & 0.97116 & 0.97644 & 0.97379 & 0.99414 & 13.44913 s& 0.95592 & 0.01937 \\ 
\hline

Densenet169 & 0.96164 & 0.94393 & 0.98554 & 0.96429 & 0.99248 & 28.31266 s& 0.90368 & 0.01386 \\ 
\hline

Densenet201 & 0.96078 & 0.94631 & 0.98103 & 0.96336 & 0.99276 & 27.89082 s& 0.90671 & 0.01558 \\ 
\hline

Darts AutoML (5 Layer) & 0.97672 & 0.97755 & 0.97815 & 0.97785 & 0.99659 & 13.15136 s& 0.95623 & \textbf{0.01171} \\ 
\hline

Darts AutoML (7 Layer) & \textbf{0.98611} & \textbf{0.98520} & \textbf{0.98842} & \textbf{0.98681} & \textbf{0.99907} & \textbf{4.01985 s} & \textbf{0.98694} & 0.01532 \\ 
\hline

\end{tabular}

}
\label{Online Result Table}
\end{table*}
\subsubsection{Neural Architecture Search}

The Darts network for the baseline data is found, with the meta network parameters set at 5 layers, 5 nodes per cell, and 5 channels per node. Due to a performance decrease when the same Darts parameters are applied to the application dataset, the Darts model for the application set is fixed at 7 layers, 5 nodes per cell, and 9 channels per node. These choices are somewhat arbitrary, but have direct impact on memory usage during the NAS and the complexity and predictive performance of the found architecture. The selections made are to allow the model to fit on a single GPU while achieving good predictive performance. The impact of these choices are discussed \cite{liu2018darts}. 

A dropout rate of 0.30 is used in the neural architecture search, the same as used in all the rest of the model training. The Darts architecture search is run for 30 epochs (approximately 13 hours), using the training data. A batch size of 96 is used, the same as the original Darts paper. The found architecture is then trained using the same hyper parameters as the models it is compared to, described next.

\subsubsection{Training Parameters}
In order to compare the performance of Darts to state-of-the-art CNNs, these models will be trained the same way as the found Darts models: \textit{Resnet18}, \textit{Resnet50}, \textit{Resnet101}, \textit{Densenet121}, \textit{Densenet169}, and \textit{Densenet201}. All the considered models share the same hyper parameters. The models are each trained for 100 epochs, use the Adam optimizer with a learning rate of 0.0005, learn on a batch size of 512, and have a dropout rate of 0.30. For each model, the epoch with the lowest validation loss is used on the test set to produce the final results for that model.  

\subsubsection{Results}

The best found normal and reduction convolutional cells structures in the baseline darts model are shown in Figures \ref{normal cell} and \ref{reduction cell}, respectively.  The two input nodes in each cell are the outputs of the previous two cells, or in the case of the first cell the duplicated output of the first layer of Darts. All the node outputs are concatenated to be the cell output. 

The training and evaluation was conducted on a VM with 14 vCPUs, 100 GB of memory, and an RTX A6000 GPU with 48 GB of VRAM.

The predictive results of the test set are shown in Table \ref{Online Result Table}. This table shows the accuracy, precision, recall, F1-score, and Area Under the Curve (AUC) for each model. Additionally, to model a real world scenario, a threshold is calculated from the validation set, such that the the validation false positive rate is 1.00\%. This models a scenario where many false positives can become overwhelming for analysts to deal with, so an effort is made to minimize them by increasing the detection threshold of the malware detection model. When the threshold is increased, this can create a delay in a positive malware detection, in real time, through false negatives at the beginning of malware execution. This is shown in the table as \textit{Delay @ Low FPR}, and is the average number of seconds elapsed before a successful detection after the malware injection point. Also shown in this section of the table is True Positive Rate \textit{TPR} and False Positive Rate \textit{FPR} at the high detection threshold (low FPR) on the test set.

Both of the Darts models that were tried are shown in the Application section of Table \ref{Online Result Table}. The first model has 5 layers, 5 nodes per cell, and 7 channels per node. The second Darts model has 7 layers, 5 nodes per cell, and 9 channels per node. 
The Darts models in both baseline and application datasets perform better in almost every area than state-of-the-art models. In the baseline set, Resnet18 and Resnet50 show better precision and recall than the Darts model, respectively. It can, however, be seen that the Darts model has a higher F1-score signifying that the Darts model better balances precision and recall on the test set better than either of the other models. The Darts model also has the lowest delay, and is under 10 seconds, meaning that most of the malware in each execution experiment was detected in the first time slice after injection. Additionally, many of the state-of-the-art models are shown to impose a significant delay in the detection of the malware, with some averaging over 2 time slices, or over 20 seconds for a successful detection. The Darts models don't always have the lowest FPR at the high detection threshold, but all results in this column are shown to be close to the 1\% target to validate the delay and TPR results.

\textit{Accuracy}, \textit{Precision}, \textit{Recall}, and  \textit{F1-Score} are shown for each model in both sets in Figures \ref{baseline_graph} and \ref{app_graph}. The average malicious prediction delay is also shown in Figures \ref{baseline_delay} and \ref{app_delay}.
The higher performance difference between the Darts models and state-of-the-art models in the application set vs the baseline set, suggests the need for AutoML derived models as data becomes more complex. Data from a server during real world use is more noisy and allows for malware execution to better hide within this noise. The neural architectures that are specifically derived based on this more complex data for this use case are more performant at identifying malware execution than generic architectures.

%% file: sections/Future_Work_Conclusion.tex
\vspace{-2mm}
\section{Future Work and Conclusion }\label{sec:future_work_conclusion}

\subsection{Future Work}
This work describes the usefulness of AutoML for malware detection. Future works can expand on the ideas of this work with different search algorithms and malware data sources, as well as create tools to even further automate the process to make layman use of these methodologies easier. 

\subsubsection{Recurrent Neural Networks}
Recurrent Neural Networks have shown near perfect results with online per-process performance metric data \cite{kimmel2021recurrent_rnn}. The Darts methodology can also be used to derive recurrent cells, and this methodology should be examined on the dataset from Section \ref{sec:online_automl} in the future.

\subsubsection{Per-Layer Granularity}
In our work in Section \ref{sec:static_automl}, once the width of the hidden layer is selected from the search space, it is fixed throughout the hidden layers of a model leading to a rectangular shape of the hidden layers in the model. However, an equivalent or a more optimal model may contain variable size layers with potentially fewer trainable parameters. A NAS process that allowed this level of granularity without an explosion of the NAS search space would prove valuable. 

\subsubsection{Refinement of Meta-Hyper-Parameters}
The set values of the meta-hyper-parameters have a significant effect on the efficacy of the AutoML process. Works such as \cite{feurer2018towards_metahp} have developed methods to optimize a search strategy within the given confines of the meta-hyper-parameters in a data driven way. Finding the appropriate bounds of these parameters, specifically tailored to the malware detection domain, is yet to be explored.

\begin{figure}[!t]
\begin{center}
\includegraphics[width=0.35\textwidth]{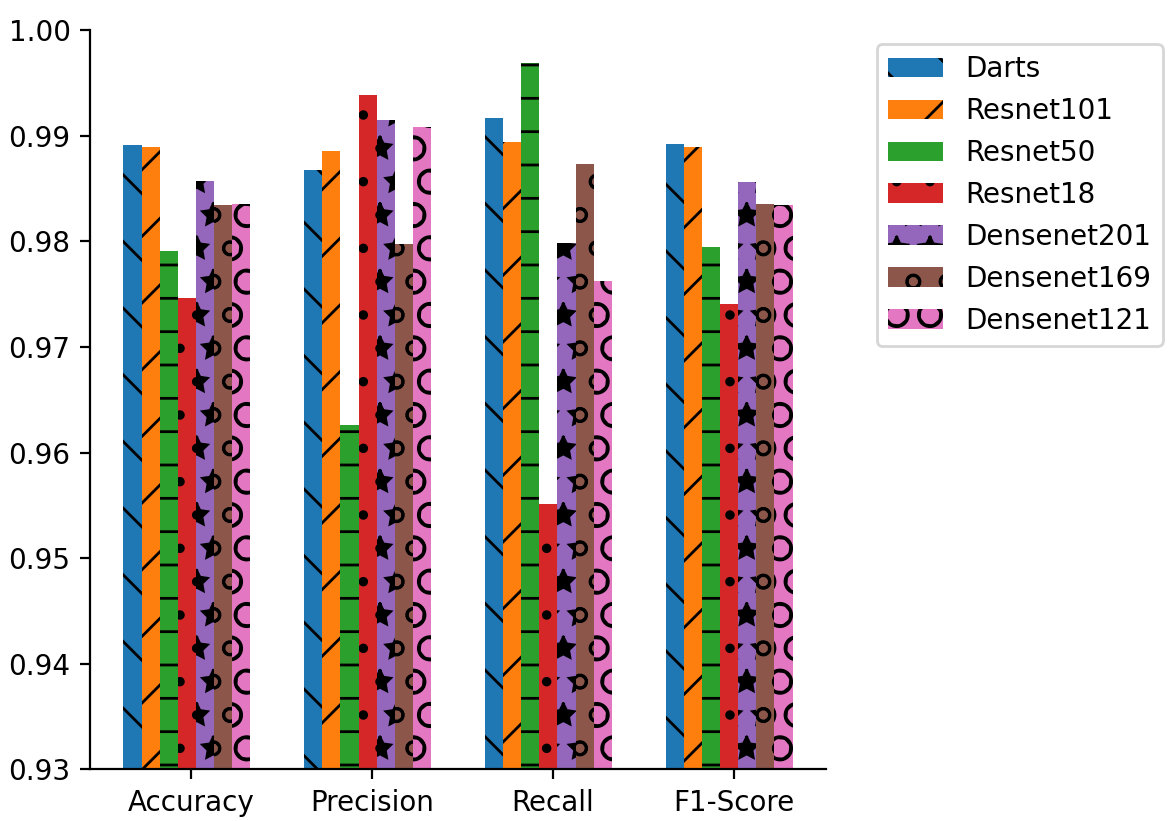}    
\end{center}
\caption{Baseline Results}
\label{baseline_graph}
\end{figure}
\begin{figure}[!t]
\begin{center}
\includegraphics[width=0.35\textwidth]{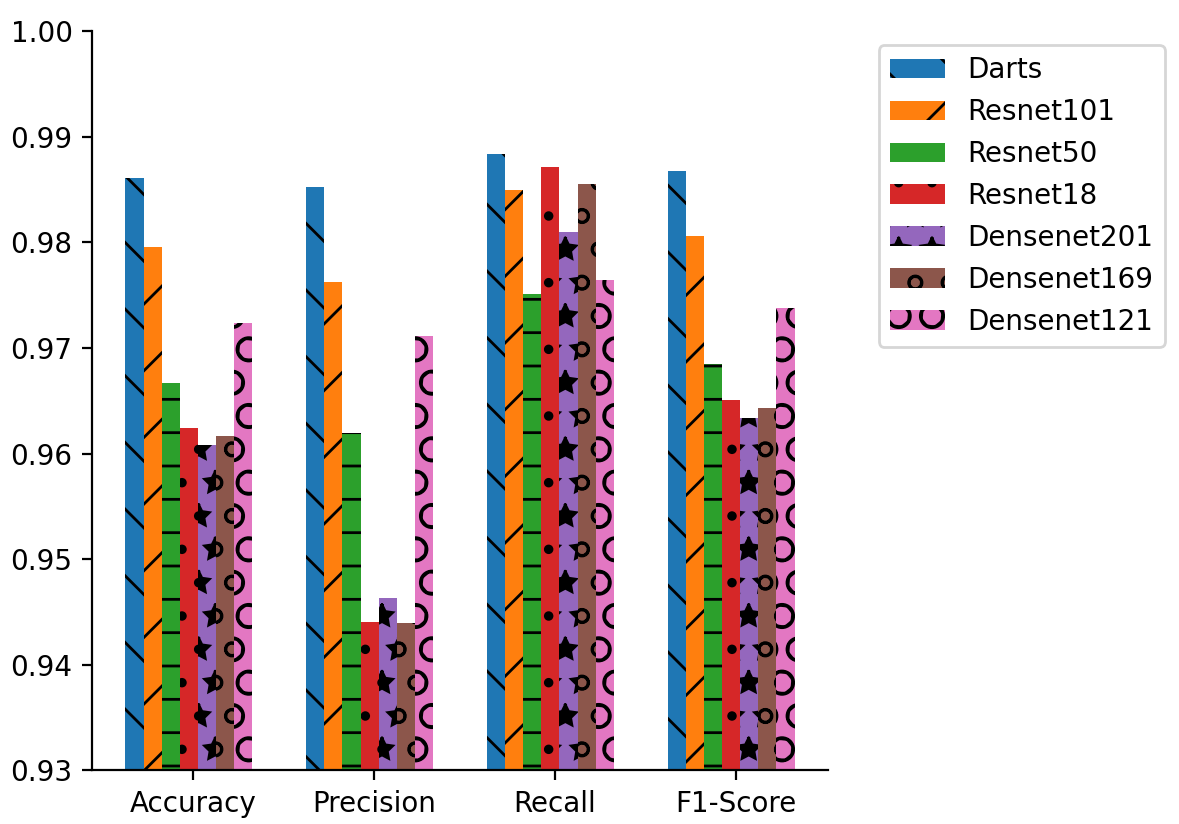}    
\end{center}
\caption{Application Results}
\label{app_graph}
\end{figure}
\begin{figure}[!t]
\begin{center}
\includegraphics[width=0.35\textwidth]{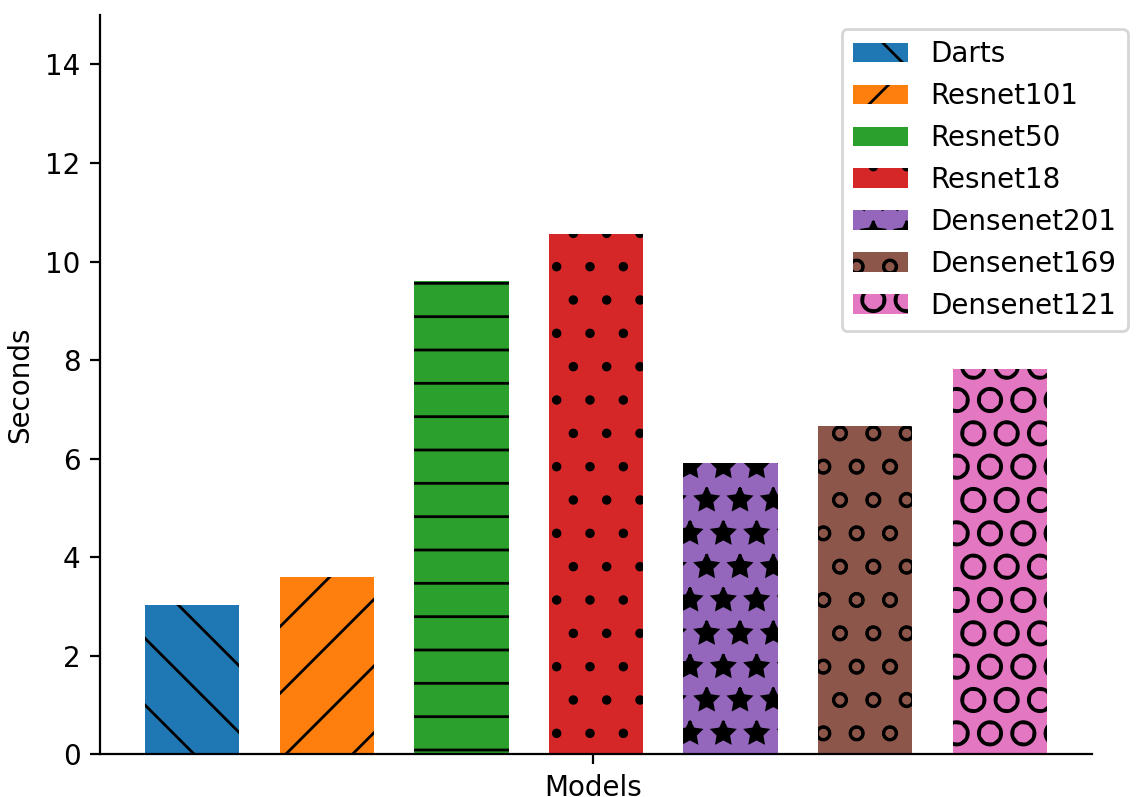}
\end{center}
\caption{Baseline Delay}
\label{baseline_delay}
\end{figure}
\begin{figure}[!t]
\begin{center}
\includegraphics[width=0.35\textwidth]{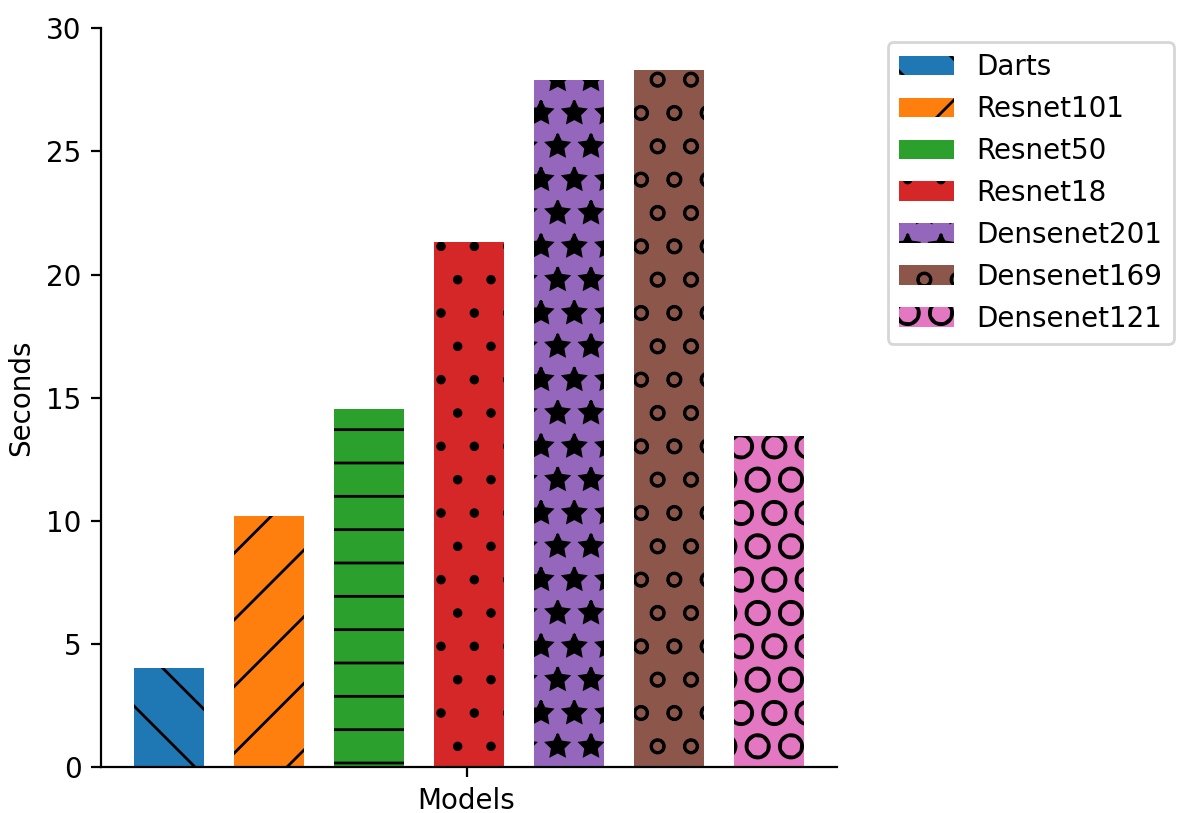}    
\end{center}
\caption{Application Delay}
\label{app_delay}
\end{figure}
Addition of auxiliary output heads to the NAS search space can also be considered meta-hyper-parameters. One of the potential labels that can be given to this data may not be of use in a strictly detection setting, but may help derive a more performant model for the required objective with auxiliary loss, just as discussed in \cite{rudd2019aloha}. Automatic inclusion of these in the search space based on label data would be valuable in automatic model searching. 
If hyper-parameter tuning is also performed as part of the AutoML process, the tuning algorithm can also be considered a meta-hyper-parameter. Depending on the evaluation metric, or rather intended performance (low false positive rate, high accuracy, etc), the found optimal parameters may differ. Algorithms such as differentiable evolution can also allow for optimization for multiple objectives (evaluation metrics). 

\subsubsection{Deep Learning Types and Ensemble Learning}
In Section \ref{sec:static_automl}, we only used FFNNs for the SOREL-20M and EMBER-2018 datasets. An analysis of using various deep learning models can be very useful. 
Further, malware data can be extracted in many forms and types of data (e.g., time series and image data). Training a machine learning model on combined dynamic time series data and statically extracted tabular data can enhance the model's detection ability. However, designing such a model can be very difficult and, as such, AutoML is the perfect candidate for this task. An AutoML system that can intelligently conform to other sources of heterogeneous data is an area for future work. 

In addition, AutoML can be utilized for ensemble learning. For instance, an AutoML system that can train multiple sub-models of different types and ensemble the sentiment of the sub-models would allow for more robust application in practice. Works such as \cite{erickson2020autogluon} ensemble many types of machine learning models, including FFNNS, to achieve better results. Extending this to other deep learning model types could prove beneficial for malware detection.

\subsubsection{User Friendly AutoML}
Designing AutoML models can be easier than designing a deep learning model from scratch, but an even more automated deep learning approach would be helpful for those with knowledge of their own data, but not necessarily deep learning. An AutoML system that could be instantiated with only training data inputs, type of data (vector, image, time-series), and primary and auxiliary labels would allow even broader access to malware detection solutions using deep learning. This framework would be able to automatically select a model type of deep learning architectures and use AutoML techniques to find a performant architecture to suit the data, making maximal use of any provided auxiliary information. Ideally, this would combine the methodologies and discussions from both Sections \ref{sec:static_automl} and \ref{sec:online_automl}. It would perfom all phases of the AutoML process efficiently, and be able to set applicable meta-hyper-parameters from details of the provided training data.

\vspace{-2mm}
\subsection{Conclusion}




In conclusion, we conjecture that Automated Machine Learning offers an effective solution for detecting malware in both static and online cloud IaaS environments. We found that AutoML generated models can perform just as well or even better than state-of-the-art models or models that have been handcrafted by experts with domain knowledge in machine learning and malware. We explored the performance of AutoML on two popular datasets static malware datasets in Section \ref{sec:static_automl}, SOREL-20M used to demonstrate efficacy on large datasets; and EMBER-2018, a dataset that was specifically curated to hinder the performance of machine learning models; with results in Tables \ref{SOREL-20M results table} and \ref{EMBER results table}. Our work on static malware datasets showed the feasibility of using AutoML as a tool for malware detection while reducing the external complexity and expertise required to train DL models.

We further explored one-shot AutoML on a new online cloud IaaS malware dataset using CNNs. Our results show that AutoML approaches can be utilized by cloud service providers and malware detection vendors to find custom deep learning models for malware detection utilizing any of a variety of data sources. The online approach we have shown can derive a custom CNN that is more capable than state-of-the-art models and contains cells that are more complex than what can feasibly be derived by hand. Importantly, we demonstrated that the difference in detection ability between AutoML models and state-of-the-art models becomes more pronounced as the noise in the input data increases, approaching the noise levels seen in real-world applications. We also elaborate on future directions to mature the use of AutoML research towards cybersecurity domains.

\vspace{-2mm}
\section*{Acknowledgements}
This work is partially funded by the National Science Foundation grants 2230609, 2043324 at Tennessee Tech University, and 2230610 at North Carolina A\&T State University.